\documentclass[12pt]{iopart}
\newcommand{\gguide}

\usepackage{graphicx}
\usepackage{iopams}
\begin{document}

\title[Excited-State $\beta$-Decays and the r-Process]{Effects of $\beta$-Decays of 
Excited-State Nuclei on the Astrophysical r-Process}
\author{M A Famiano}
\ead{michael.famiano@wmich.edu}
\address{Department of Physics,
Western Michigan University, 
Kalamazoo, MI 49008-5252 USA}
\author{R N Boyd}
\ead{boyd11@llnl.gov}
\address{Lawrence Livermore National Laboratory, Livermore, CA 94550 USA}
\author{T Kajino}
\ead{kajino@nao.ac.jp}
\address{Department of Astronomy, University of Tokyo, 7-3-1 Hongo,
Bunkyo-ku, Tokyo 113-0033 Japan}
\address{Department of Astronomical Science, Graduate University for
Advanced Studies, 2-21-1 Osawa, Mitaka, Tokyo 181-8588 Japan}
\address{National Astronomical Observatory, 2-21-1 Osawa, Mitaka, Tokyo 
181-8588 Japan}
\author{ K Otsuki}
\ead{otsuki@uchicago.edu}
\address{Department of Physics and Astronomy, The University of Chicago, Chicago, IL 60637 USA}
\author{M Terasawa}
\ead{mariko@cns.s.u-tokyo.ac.jp}
\address{Center for Nuclear Study, The University of Tokyo, 2-1 Hirosawa, Wako, Saitama 351-0198 Japan}
\author{G J Mathews}
\ead{gmathews@nd.edu}
\address{Department of Physics, University of Notre Dame, 225 
Nieuwland Science Hall, Notre Dame, IN 46556-5670 USA}

\begin{abstract}
A rudimentary calculation is employed to evaluate the possible effects of $\beta$-decays of excited state
nuclei on the astrophysical r-process.
Single particle levels calculated with the FRDM are adapted to the calculation of $\beta$-decay 
rates of these
excited state nuclei.  Quantum numbers are determined based on proximity to Nilson model levels.   
The resulting rates are used in an r-process network calculation in which a supernova hot-bubble model is coupled
to an extensive network calculation including all nuclei between the valley
of stability and the neutron drip line and with masses 1$\le$A$\le$283.  
$\beta$-decay rates are included as functional forms of the 
environmental temperature.  
While the decay rate model used is simple and phenomenological, it is consistent across all 3700 nuclei 
involved in the r-process network calculation.  This represents an approximate first estimate 
to gauge the possible effects of excited-state $\beta$-decays on r-process freezeout abundances.
\end{abstract}

\pacs{26.30.Hj,26.50.+x}

\section{Introduction}
\label{Introduction}
The r-process is responsible for the synthesis of roughly half of all nuclei
heavier than A$\sim$70 and all of the actinides 
\cite{cowan90,wallerstein97}. 
The solar system r-process abundances act as the canonical 
constraint to r-process theories as well as the prime indicator of the 
success of r-process models.
Several
r-process sites have been proposed; the hot-bubble region of a 
type II supernova (SNII) has been modeled fairly successfully.
The composition of the environment in which the r-process occurs might be 
expected to have a 
profound effect on the final abundance distribution.   
Observations indicate that the r-process site is primary \cite{sneden98}
and further
evidence may suggest that the r-process is also 
unique \cite{cowan99}; it may occur in a single
site or event.  The uniqueness of the r-process site, however, remains
a subject of study \cite{qian98}.

Nuclear properties also constrain the r-process,
and the purpose of this work is the examination of
one particular characteristic - $\beta$-decay - as it relates to the 
r-process.  The $\beta$-decay inputs, and other nuclear physics inputs, have
been shown \cite{woosley94,meyer95} to have important effects on the success
or failure of r-process models.  This is somewhat unfortunate, as properties 
of only a few nuclei
on the neutron closed
shells closer to stability have been experimentally determined, while data
for the rest are relegated to calculation \cite{cowan90}.  Of paramount
importance is the determination of nuclear masses and $\beta$-decay rates.
Nuclear mass formulae based on the microscopic properties of nuclei are
slowly replacing the empirical droplet models, and these can change
resulting reaction rates by factors as large as 10$^8$ \cite{goriely96}.  As
well, the r-process path is affected by the choice of mass formula,
since the path roughly follows a line of constant S$_n$ \cite{cowan90}.  

A successful r-process calculation must predict an abundance peak for nuclei in the 
A$\sim$195 region - a difficulty over much of the r-process parameter space.  
However, as discussed below, $\beta$-decays from excited state nuclei may help to
mitigate this difficulty.  They do so by allowing the r-process to proceed at 
a faster rate, thereby enhancing the abundances at higher
masses. 

For the purposes of this study, the most recent semi-gross theory of 
$\beta$-decay \cite{nakata97} has been adapted to neutron-rich nuclei relevant to the
r-process.  The ability of this model to determine decay properties of an
extremely wide range of nuclei with reasonable accuracy and speed makes it ideal for this
preliminary calculation.  In particular, the semi-gross theory has good
agreement for very neutron-rich nuclei \cite{ichikawa05,asai99}.  It has also been used
to improve the accuracy of decay rates for astrophysical calculations by incorporating
first-forbidden transition strengths \cite{moller03}.
In its original form, the gross theory of $\beta$-decay assumed that the
energy states of a
nucleus consist of a smoothed distribution with transition
strengths that peak at or near the energy of the isobaric analog state
\cite{takahashi72,takahashi69}.  Subsequent evolutions of the gross theory incorporated
strength functions allowing for transitions of higher forbiddenness \cite
{takahashi71}, as well as improvements over the original theory to include
odd-odd effects \cite{nakata95}, sum rules \cite{tachibana90}, even-odd mass differences
\cite{koyama70}, and improvements on the strength functions \cite
{kondoh85}. 

While the precision of the Fermi gas model used is consistent with that of the 
gross theory itself, more accurate models might be used.  Recently, shell
effects in the parent nucleons have been taken into account by using 
an energy distribution for single-particle states \cite{nakata97,kondoh76};
this is denoted as the ``semi-gross theory.''
In these models, the energy distribution of the daughter states is still 
assumed to be smooth. Thus, the transition strength functions depend on the
quantum numbers only of the initial parent states and are independent of the
energy of the parent state; transition types are based on a statistical weight
for a particular parent state to make a particular type of transition.  The 
advantage of using single-particle strength 
functions is that quantum numbers can be assigned to the states easily, lending
a better notion of the actual strength of each type of transition involved.

Since decay rates of 
nearly all of the nuclei along the r-process path have yet
to be studied in a laboratory, r-process calculations rely heavily on calculated
decay rates. Further, the temperature of the r-process environment (~10$^9$K)
necessitates accounting for nuclei in excited states, especially given the
expected high level density of these far-from-stability nuclei. Some of the
effects that might be expected if one considers excited state nuclei in
r-process simulations include increased (n,$\gamma$) and ($\gamma$,n) rates,
which might shift the r-process path, but would tend to counteract each
other as the r-process is generally presumed to proceed at 
(n,$\gamma$)$\leftrightarrow$($\gamma$,n) equilibrium,
increased neutrino spallation rates, tending to enhance smoothing in
post-processing, and increased beta-decay rates.  However, at signicant
excitation, neutron separation energies are low enough that neutron emission may
be a dominant decay mode.  This is discussed briefly along with the discussion of
excited-state decays in the network calculation.

Section \ref{grosstheory} of this paper is an overview of $\beta$-decay calculations and average properties
in calculating decay rates calculated using the more recent semi-gross theory.  
This includes a brief review of the semi-gross theory of $\beta$-decay and a description of the calculations of
single-particle states and their relationship to $\beta$-decay in \S\ref{spstates}.  Energy levels are
calculated in \S\ref{spcalc} using the Finite Range Droplet Model (FRDM).  Results of
$\beta$-decay calculations of excitated state nuclei with the adapted FRDM are discussed in \S\ref{results},
along with their potential astrophysical importance.  Application of these results to an r-process network calculation 
is made in \S\ref{r-process} with results from a model with several environmental parameter sets described in 
\S\ref{network_results}.
Future work and possibilities are discussed in \S\ref{conclusion}, along with experimental possibilities.

\section{Average Nuclear Properties in $\beta$-Decay}
\label{grosstheory}
Average theories of $\beta$-decay, such as the gross theory and the more recent semi-gross theory,
make 
them useful in estimating and adapting decay rates of several nuclei to large systems such as 
astrophysical network calculations.
The major assumption in these theories is that nuclear level
densities are high enough to make integration over states a fairly accurate
approximation by replacing the matrix elements with transition 
probability
distribution functions.  For example, the decay rate for 
Fermi and GT transitions, as calculated in the gross theory by
Takahashi and Yamada, is given by \cite{takahashi69}:
\begin{equation}
\label{eq1}
\lambda = \frac{m^{5}_{e}c^{4}}{2 \pi \hbar^{7}} \int_{-Q}^0 \int_{\varepsilon
_0(E)}^{\varepsilon_1} \sum_\Omega D_\Omega(E,\varepsilon) f_\Omega (-E)
\frac{dN_1}{d\varepsilon} W(E,\varepsilon) \, d \varepsilon \, dE
\end{equation}
which integrates over the product of the transition probability function
$D_{\Omega}(E,{\varepsilon})$ between a parent nucleus state at 
energy $\varepsilon$ and a daughter state with a transition energy $E$, the
density of states $dN_1/d\varepsilon$ of the decaying nuclei; the weighting 
function of the final state nucleus W(E,$\varepsilon$), which imposes
the Pauli exclusion principle on the rate calculation; and the 
form factor
f$_\Omega$(-E), which is the product of the emitted electron wave functions 
and the
Fermi function for a transition of type $\Omega$. 
Terms used in the
equations of this paper are summarized in Tables \ref{terms} - \ref{termsb}.

The density of parent states $\frac{dN}{dE}$ supplies information on the nucleons
available for decay, while the weighting function controls the availability of the
daughter
states to which the nucleons can decay.  In the case of an even number of daughter nucleons, the 
available 
energy levels are filled up to a value $\varepsilon_1$-Q, and in the case of 
an odd number, there is a hole at the highest energy in the ground state. 
Therefore, the weighting function is given by \cite{takahashi69}:
\begin{equation}
\label{eq5}
\begin{array}{l@{\quad}l}
{W(E,\varepsilon)}_{even} = \left\{ \begin{array}{l@{\qquad \qquad}l}
1 & \varepsilon + E \ge \varepsilon_1 - Q \\
0 & \varepsilon + E < \varepsilon_1 - Q
\end{array} \right. & \\
~ & \\
{W(E,\varepsilon)}_{odd~} =\left\{ \begin{array}{l@{\qquad \qquad}l}
1 & \varepsilon + E \ge \varepsilon_1 - Q + 2 \Delta \\
\frac{1}{n_3} & \varepsilon + E = \varepsilon_1 - Q \\
0 & \mbox{otherwise}
\end{array} \right.
\end{array}
\end{equation} 
The value of n$_3$, the number of daughter nucleons promoted to the highest 
energy due to the pairing forces, is:
\begin{equation}
n_3 = \int_{\varepsilon_1 - Q}^{\varepsilon_1 - Q + \Delta} \left(
\frac{dN_2}{d\varepsilon} \right) \, d\varepsilon
\end{equation}
where dN$_2$/d$\varepsilon$ is the density of states of protons (neutrons) 
in $\beta^{-(+)}$ decay and $\Delta$ is the pairing gap.

The decay must be energetically possible, so the lowest possible energy that
a decaying nucleon can have is $\varepsilon_1$-Q-E, recalling that E is the
(negative) transition energy of the decaying nucleon.  Thus, 
$\varepsilon_0$ in Equation
\ref{eq1} equals $\varepsilon_1$-Q-E, removing that 
variable from the equation.
\subsection{Single-Particle States}
\label{spstates}
Although a smooth model is adaptable to decays, particularly from the ground-state, excited 
states present the added difficulty of potentially altering the strength functions.  For an 
accurate treatment, an estimate of the quantum numbers of these states is made.
In the initial calculation, discrete energy levels are used.  Thus, the density of states for
neutrons (protons) in $\beta^{-(+)}$ becomes:
\begin{equation}
\label{eq7}
\left(\frac{dN}{d\varepsilon}\right)_{N/Z}=\sum_i n_i \delta(\varepsilon-\varepsilon_i)
\end{equation}
where the value of n$_i$ corresponds to the number of nucleons in each
state i.  For discrete Fermi states, the levels are two-particle levels with
n=zero, one, or two.  The weighting function
W(E,$\varepsilon$) in Equation \ref{eq1} can also be adjusted for single 
particle daughter levels.  In the original gross theory, the weighting 
function takes on values between zero and one and is typically zero or one 
except at a few special points.  If energy levels are discrete, then the
value of the weighting function for a two particle level is either zero, one,
or 0.5 for a level that is already filled, empty, or half-filled (containing
a single nucleon) respectively.  In this formulation, the weighting function 
corresponding to the availability of the proton (neutron) level structure of 
the daughter nucleus,
in $\beta^{-(+)}$ depends on the energies of both the daughter nucleon levels 
$\varepsilon_k$
and the parent nucleon levels $\varepsilon_i$, because their difference is
necessary to determine if the decay is energetically possible:
\begin{equation}
\label{eq8}
W(\varepsilon_i,\varepsilon_k)=n_k\theta(\varepsilon_i-\varepsilon_k\pm
\Delta_{NP}-m_e)
\end{equation}
where the $\pm$ is for $\beta^\mp$-decay, m$_e$ is the electron mass, and
$\Delta_{NP}$ is the neutron-proton mass difference.

Each composite state is treated as a mixture of quantum states that will
depend on the model used.  Therefore, the mixture in a level i is
characterized by a sum of eigenstates weighted by the coefficient 
$\omega$(i;n,$l$,$j$) as
a function of the quantum numbers of n, $l$, $j$.  For this study, the quantum
numbers are for eigenstates of a spherical shell model based on a Woods-Saxon 
potential.  The coefficients for a level i are normalized to unity:
\begin{equation}
\label{eq9}
\sum_{n,l,j}\omega(i;n,l,j)=1
\end{equation}
for each level i.  The description and calculation of weights, as well as the
determination of single-particle states,  are further discussed in \S 
\ref{spcalc}.

Transition strength functions are then  adjusted to 
accomodate the single particle states in the daughter nucleus.
Discrete functions are used to specify the most probable 
location and quantum numbers of an available state.
Then, the transition strength for a specific level i can be written as a sum
of the product of weights for the parent and daughter levels, since levels
in the daughter nucleus are now assigned.  The transition strength
for favored decays now becomes:
\begin{equation}
\label{eq9a}
D_\Omega(E,\varepsilon)=D_\Omega(\varepsilon_f,\varepsilon_i)=\sum_{\zeta,
\xi}\omega(i,\zeta)\omega(f,\xi)\Lambda_\Omega(\zeta,\xi)
\end{equation}
where the function $\Lambda$ is used to designate the selection rules (for
the set of quantum numbers $\zeta$ and $\xi$ of the initial state i and final 
state f respectively) which satisfy the transition type $\Omega$:
\begin{eqnarray}
\label{eq9aa}
\Lambda_F & = & \delta\left(j_f,j_i\right)\delta\left(l_f,l_i\right)
\delta\left(T_{0,f},T_{0,i}\mp{1}\right)
\nonumber\\
\Lambda_{GT} & = & \delta\left(j_f,j_i\right)\delta\left(l_f,l_i\right)
\delta\left(\tau_{0,f},\tau_{0,i}\mp{1}\right)\nonumber\\
~&~&+\delta\left(j_f,j_i\pm{1}\right)
\delta\left(l_f,l_i\right)
\delta\left(\tau_{0,f},\tau_{0,i}\mp{1}\right)
\nonumber\\
\Lambda_{1,V} & = & \delta\left(j_f,j_i\pm{1}\right)\delta\left(l_f,l_i\pm{1}\right)
\delta\left(\tau_{0,f},\tau_{0,i}\mp{1}\right)
\nonumber\\
\Lambda_{1,A} & = & \delta\left(j_f,j_i\right)\delta\left(l_f,l_i\pm{1}\right)
\delta\left(\tau_{0,f},\tau_{0,i}\mp{1}\right)
\nonumber\\
& & +\delta\left(j_f,j_i\pm{1}\right)
\delta\left(l_f,l_i\pm{1}\right)\delta\left(\tau_{0,f},\tau_{0,i}\mp{1}\right)
\nonumber\\
& & +\delta\left(j_f,j_i\pm{2}\right)
\delta\left(l_f,l_i\pm{1}\right)\delta\left(\tau_{0,f},\tau_{0,i}\mp{1}\right)
\end{eqnarray}

The strength
functions are also normalized to satisfy the necessary sum rules \cite
{takahashi73}.  For a single energy level, the Fermi and GT strength functions
are normalized to unity:
\begin{equation}
\label{eq9b}
\int^\infty_{-\infty}D_{F,GT}(E,\varepsilon)dE\rightarrow\sum_fD_{F,GT}
(\varepsilon_f,\varepsilon_i)=1
\end{equation}
while the vector and axial first-forbidden functions are normalized 
as \cite{takahashi73}:
\begin{equation}
\sum_fD_{V,A}(\varepsilon_f,\varepsilon_i)=\frac{R^2}{5}
\end{equation}
where R is the nuclear radius. 
By substituting into Equation \ref{eq1}, the decay rate for $\beta^\mp$ decay
is now:
\begin{eqnarray}
\label{eq10}
\lambda & = & \frac{m_e^5c^4}{C 2\pi^3\hbar^7}\left\{\sum_i \sum_k n_i n_k 
\theta
(\varepsilon_i-\varepsilon_k\pm{\Delta}_{NP}-m_e)\right.
\nonumber\\
& & \times\left.\sum_\zeta \sum_\xi 
\sum_\Omega \omega(i;\zeta) \omega(k;\xi) \Lambda_\Omega(\zeta,\xi) 
\left|G_\Omega\right|^2 \left(2l_\Omega+1\right) f_\Omega(E)\right\}
\end{eqnarray}  
The variables $\zeta$, $\xi$, and $\Lambda_\Omega(\zeta,\xi)$ represent the 
quantum numbers associated with
the states comprising the parent and daughter states respectively and the 
factor to satisfy the necessary selection rules, as mentioned above.  The decay order
is given by $l$.  The
summation over $\Omega$ is simply the sum over all possible transition orders in the
decay.  G$_\Omega$ is the appropriate coupling constant (vector or 
axial-vector) for
the transition.  The value of C is
equal to 1 for allowed transitions and ($\hbar$/m$_e$c)$^2$ for first forbidden
transitions, the only possibilities considered
in this paper.  The value E is the transition energy, and
is simply the difference in parent and daughter energy states plus (minus) the
neutron-proton mass difference and the electron mass in $\beta^{-(+)}$ decay. 

\begin{figure}
\includegraphics[angle=90,width=15cm]{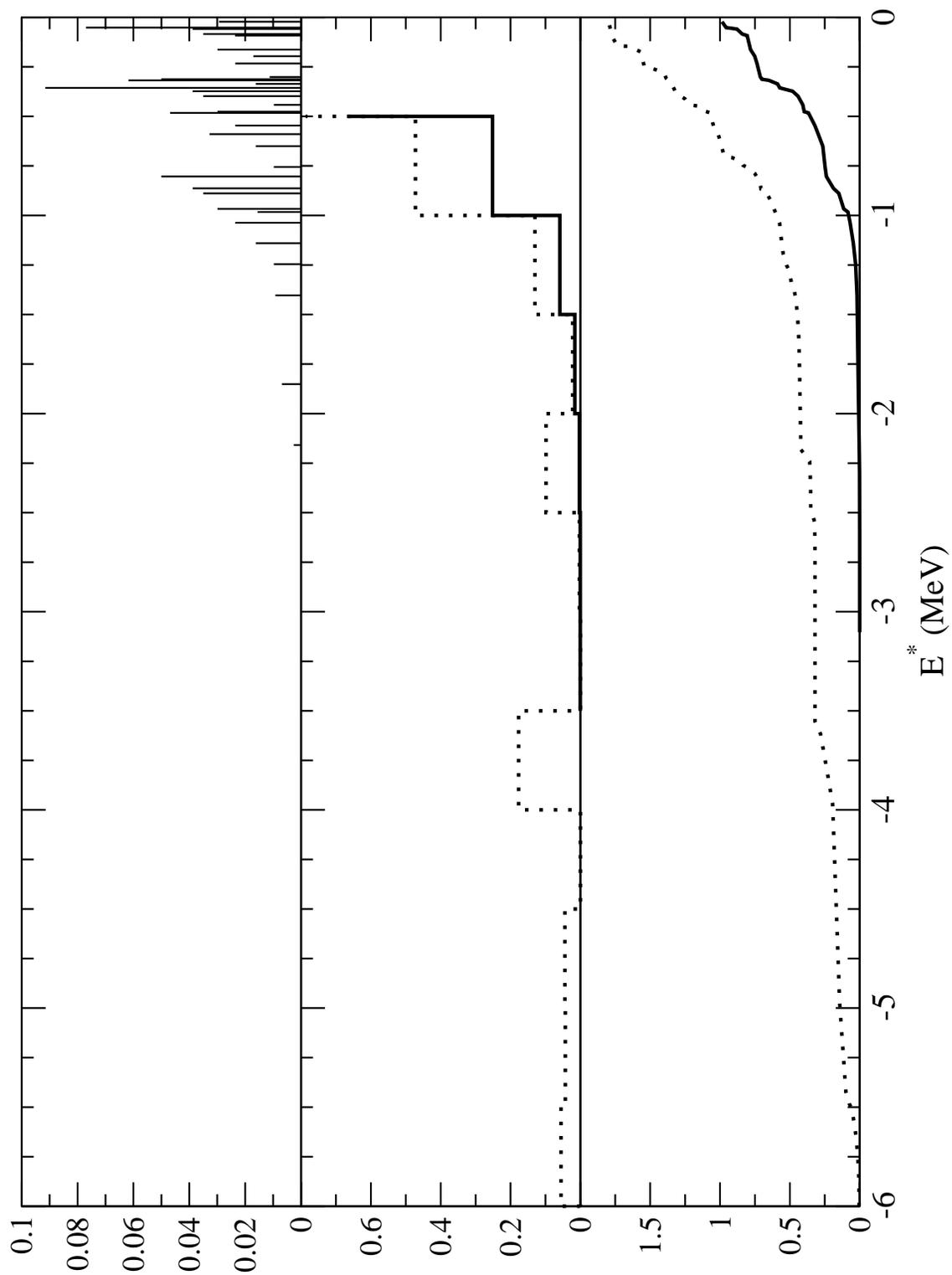}
\caption{\label{gtstr} Transition strength function $\left|M_{GT}(E)\right|^2$ 
assuming discrete levels for $^{132}$Sn.  The top graph is the discrete transition strength
function.  The middle graph is a histogram of the discrete function with 0.5 MeV bins, and the
bottom graph is the integrated function.  In the lower two graphs, the solid line corresponds to the
ground-state of $^{132}$Sn, and the dotted line corresponds to the first excited state.}
\end{figure}

\begin{figure}
\includegraphics[angle=90,width=16cm]{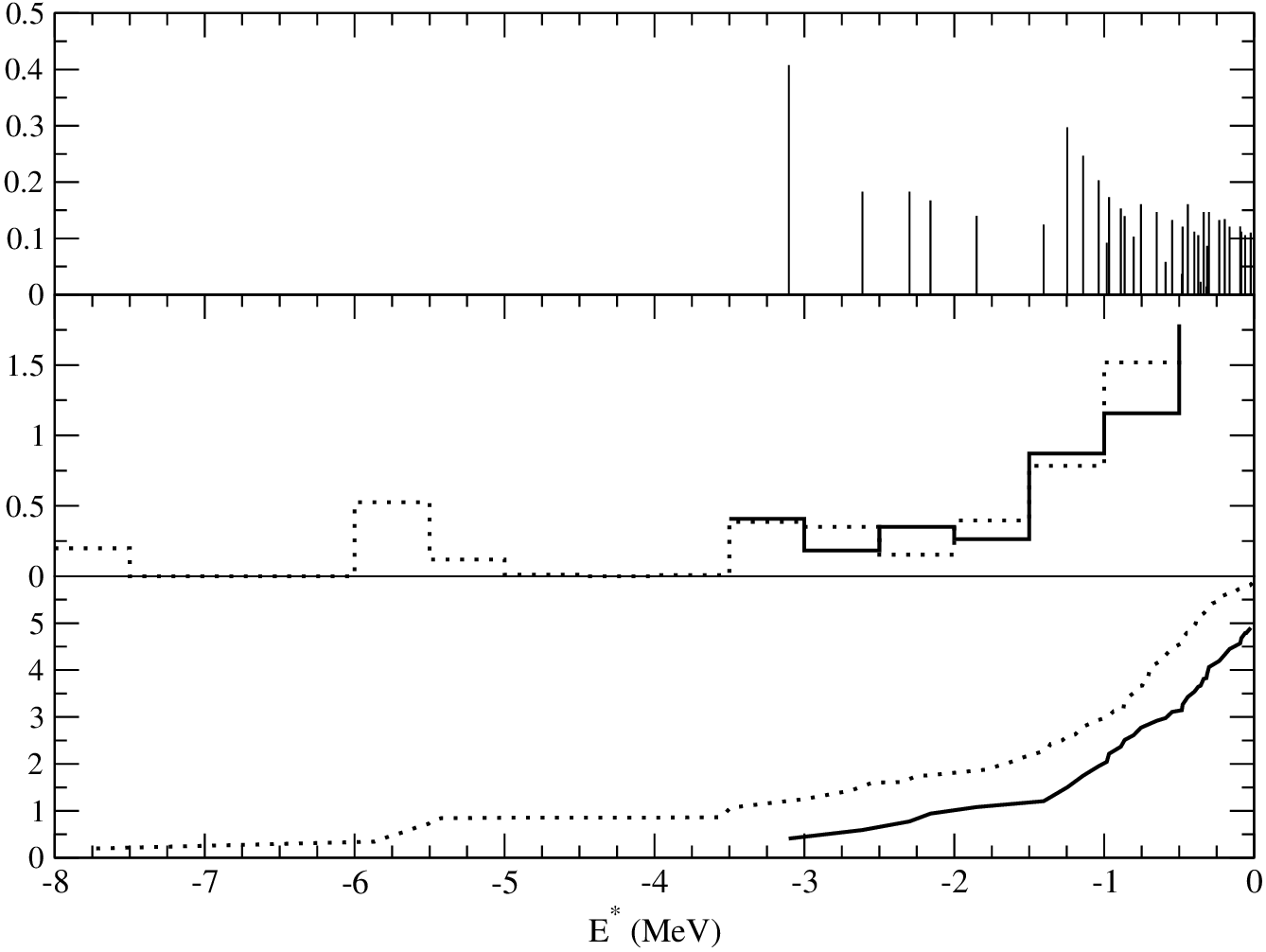}
\caption{\label{axstr} Transition strength function $\left|M_A(E)\right|^2$ assuming discrete 
levels for $^{132}$Sn.  See Figure \ref{gtstr} for an explanation
of the graphs.} 
\end{figure}

\begin{figure}
\includegraphics[angle=90,width=16cm]{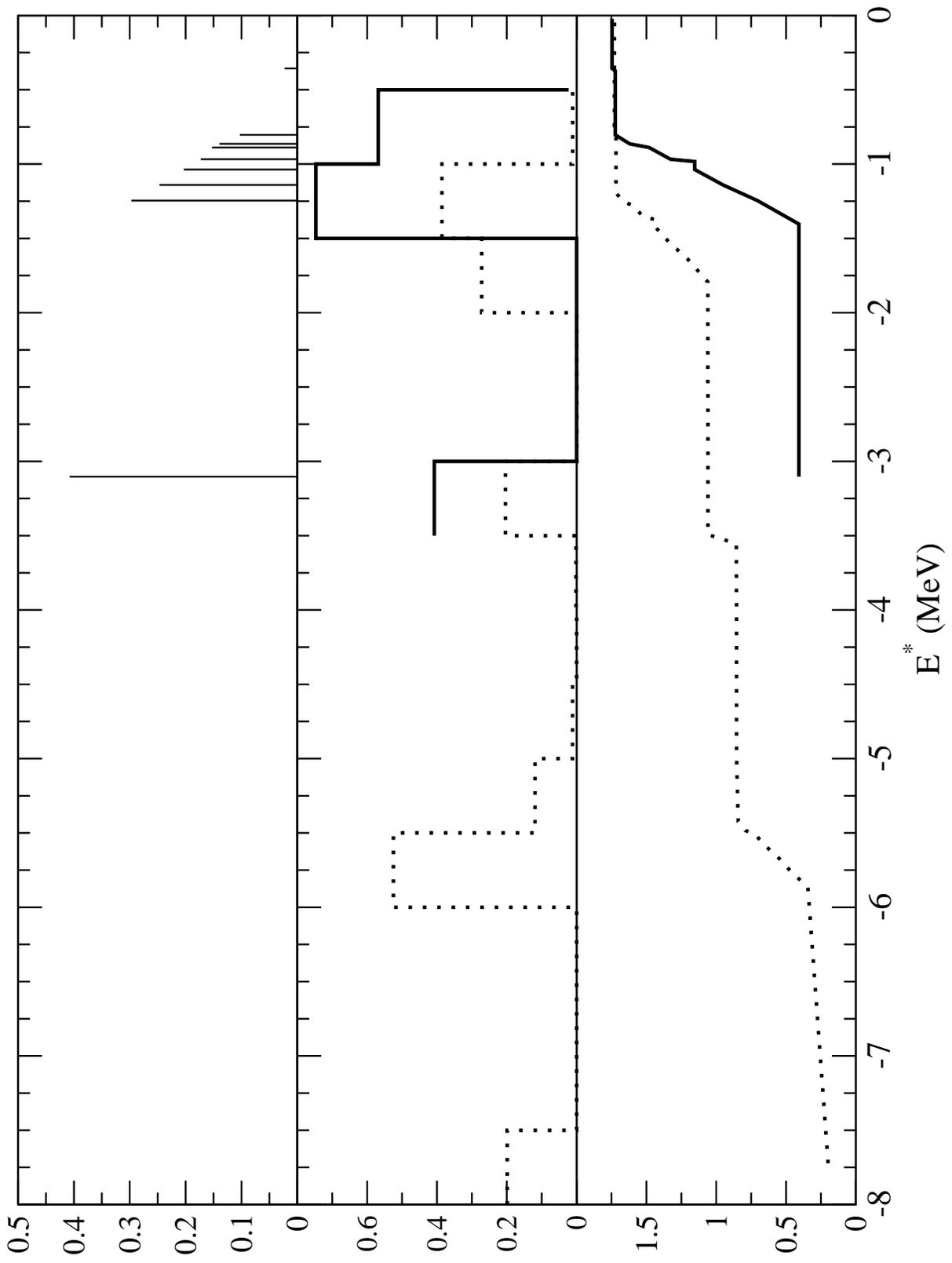}
\caption{\label{vecstr} Transition strength function $\left|M_V(E)\right|^2$ 
assuming discrete levels $^{132}$Sn.   See Figure \ref{gtstr} for an explanation
of the graphs.} 
\end{figure}

The transition matrix elements for a nucleus are then calculated as:
\begin{equation}
\label{eq11}
\begin{array}{l@{\quad}l}
\left|M_\Omega(E)\right|^2 & =\left|M_\Omega(\varepsilon_i-\varepsilon_k)\right|^2 \\
 & =\sum_i\sum_k\sum_\zeta \sum_\xi n_i n_k \omega(i;\zeta)
\omega(k;\xi)\Lambda_\Omega(\zeta,\xi)\\
\end{array}
\end{equation}
The daughter level k is determined by
the parent level i, the transition energy E, and the transition order $\Omega$.  
The sum over quantum states
$\zeta$ in the i$^{th}$ parent level as well as the sum over states $\xi$ 
in the j$^{th}$ daughter level 
are both necessary.  Though the daughter state is constrained by the 
transition selection rules for a transition $\Omega$, there may be several 
quantum states so allowed, so the sum over 
states $\xi$ is necessary.
As an example, the discrete transition strength matrix elements for $^{132}$Sn 
are shown in Figures \ref{gtstr} - \ref{vecstr}.  One should note the 
characteristic broad peak corresponding to the GT matrix element (as 
compared to the Fermi matrix element, which has no viable transition strengths relevant
for $\beta^{-}$ decay), as well as the double peak
characteristic of the first-order transition elements.

It can be seen from Equation \ref{eq10} that excited states are completely
accounted for in the values n$_i$ and n$_k$.  This makes the utility of the
equation obvious.  As well, nuclear excitations can take the form of 
proton or neutron single-particle excitations without regard to the type
of decay.  The inclusion of selection rules is accomplished through a specification 
of quantum numbers of individual states.  Furthermore, if all of the possible energy 
states
above the Fermi surface are specified, the population of these states are assumed to
be in thermodynamic equilibrium, and the partition function
is then determined.  Therefore, at temperatures peculiar to the 
r-process environment (or any environment), the values n$_i$ and n$_k$ can be
substituted with the probability functions for a statistical ensemble in
thermodynamic equilibrium.
\subsubsection{Form Factors}
\label{forms}
Besides the calculation of single particle energies, the only remaining
term in the rate equation is the f-function which, based on the
order $\Delta{l}$, rank $\Delta{j}$, and vector nature of the transition 
(axial vector or vector) for
$\beta^\mp$-decay is:
\begin{equation}
\label{eq18}
f_{l,j,A/V}(E)=\int_1^{W_0}F_{l,A/V}S_{l,j,A/V}^{\mp}pWq^2dW
\end{equation}
where F$_{l,A/V}$ is the Fermi function, which is taken to be the same in all
cases; F$_{l,A/V}$=F$_0$, independent of $\Delta{l}$.  The quantity 
S$_{l,j,A/V}$ is the factor for the Coulomb interaction between the outgoing
electron wave function and the
nucleon wave functions.  W is defined as the electron kinetic energy 
E$_e$/(m$_e$c$^2$)$\pm$1 for $\beta^\mp$-decay, and W$_0$ corresponds to the 
maximum possible electron energy E for a transition.  Also, the terms p and q are given by 
p=(W$^2$-1)$^{1/2}$
and q=W$_0$-W.  The forms of the wave function factors used are those of 
reference \cite{takahashi71}.
\subsection{Calculation of Single-Particle States}
\label{spcalc}
\subsubsection{Energy Levels of the Single-Particle States}
The finite range droplet model (FRDM) with the Lipkin-Nogami (LN) pairing force is used to calculate
single-particle levels for the various nuclei.  The formula is referred to as a 
microscopic-macroscopic - or ``mic-mac'' model.  The macroscopic part is a smooth function of proton
number Z and neutron number N based on a finite range liquid drop model.  The microcopic part
is used to calculate individual shell energies and is based on a folded-Yukawa single-particle potential.    
The macroscopic part is described if reference \cite{moller00}.  
It is based on the overall charge,
mass, and the nuclear shape.  The parametrized macroscopic part, as described in the reference is:
\begin{equation}
\label{frdmeqn}
\begin{array}{l@{\quad}l}
\fl E_{mac}(Z,N,shape)=\\
\fl +\Delta{M}_HZ+\Delta{M}_nN & \mbox{mass excess of hydrogen and neutron}\\
\fl +\left({-a_1+J\bar{\delta}^2-\frac{1}{2}K\bar{\epsilon}^2}\right)A & \mbox{volume energy}\\
\fl +\left({a_2+\frac{9}{4}\frac{J^2}{Q}\bar{\delta}^2\frac{B_s^2}{B_1}}\right)A^{2/3}&
\mbox{surface energy}\\
\fl + a_3A^{1/3}B_k & \mbox{curvature energy}\\
\fl + a_0A^0 & \mbox{A$^0$ energy}\\
\fl + c_1 \frac{Z^2}{A^{1/3}}B_3 & \mbox{Coulomb energy}\\
\fl - c_2Z^2A^{1/3}B_r & \mbox{volume redistribution energy}\\
\fl - c_4\frac{Z^{4/3}}{A^{1/3}} & \mbox{Coulomb exchange correction}\\
\fl - c_5Z^2\frac{B_wB_s}{B_1} & \mbox{surface redistribution energy}\\
\fl + f_0\frac{Z^2}{A} & \mbox{proton form factor correction}\\
\fl ~&\mbox{to Coulomb energy}\\
\fl - c_a\left({N-Z}\right) & \mbox{charge asymmetry term}\\
\fl + W\left({\left|I\right| + 
\left\{\begin{array}{l@{\quad}l}
1/A & \mbox{Z=N=odd}\\
0 & \mbox{otherwise}\end{array}\right.}\right) & \mbox{Wigner energy}\\
\fl + \left\{\begin{array}{l@{\quad}l}
\bar{\Delta}_p+\bar{\Delta}n-\delta_{np} & \mbox{Z and N odd}\\
\bar{\Delta}_p & \mbox{Z odd, N even}\\
\bar{\Delta}_n & \mbox{Z even, N odd}\\
0 & \mbox{Z and N even}\end{array}\right. & \mbox{average pairing energy}\\
\fl - a_{el}Z^{2.39} & \mbox{energy of bound electrons} 
\end{array}
\end{equation} 
The parameters for this expression, as well as their derivations, are described extensively in 
\cite{moller00}, and will be described briefly here.  Many of the constants used in this relationship are described in 
Table \ref{frdm_param}.  Other derived quantities are given in the table, as well.
The standard droplet model defines the shape-dependent quantities \cite{myers74} as the surface energy $B_s$,
the surface redistribution energy $B_w$, the curvature energy $B_k$, and the volume redistribution energy 
$B_r$.  
These are solved analytically by integrating the appropriate quantities over the surface of the nuclear 
volume:
\begin{equation}
\label{ldm_eqns}
\begin{array}{l@{\quad}l@{\quad}l}
B_s & = &  \frac{1}{4{\pi}r_0^2A^{2/3}}\int_S{dS}\\
B_w & = & \frac{225}{64{\pi}^3r_0^6A^2}\int_S{\left[\tilde{W}\left(\vec{r}\right)\right]^2dS} \\
B_k & = & \frac{1}{8{\pi}r_0A^{1/3}}\int_S{\left({\frac{1}{R_1} - \frac{1}{R_2}}\right)dS} \\
B_r & = & \frac{1575}{64{\pi}^3r_0^7A^{7/3}}\int_V{\left[\tilde{W}\left(\vec{r}\right)\right]d^3r}\\
\end{array}
\end{equation}
where $R_1$ and $R_2$ are principal radii of curvature 
(two principal radii assume deformations having no higher
than quadrupole terms in the macroscopic model).  The volume term is specified as
$\tilde{W}\left(\vec{r}\right)=W(r)-\tilde{W}$ and:
\begin{equation}
\label{ldm_volterms}
\begin{array}{l@{\quad}l@{\quad}l}
W\left(\vec{r}\right) & = & \int_V{\frac{1}{\left|{\vec{r} - \vec{r}\prime}\right|}d^3r^{\prime}}\\
\tilde{W} & = & \frac{3}{4{\pi}r_0^3A}\int_V{W\left({\vec{r}}\right)d^3r}
\end{array}
\end{equation}
The quantities $B_1$ and $B_3$ are the relative generalized surface and Coulomb energies respectively for a 
deformed shape of volume $V=\frac{4\pi}{3}\left({r_0A}\right)^3$.  These quantities are developed to treat nuclei
under finite compressibility and diffuseness $a_{den}$.  The generalized forms of these quantities are:
\begin{equation}
\label{frdm_shapes}
\begin{array}{l@{\quad}l}
B_1&= \frac{1}{8\pi^2r_0^2a^4}\int\int_V{\left({2-\frac{\left|\vec{r}-\vec{r}^\prime\right|}{a}}\right)
\frac{e^{\left|\vec{r}-\vec{r}^\prime\right|}/a}{\left|\vec{r}-\vec{r}^\prime\right|/a}
d^3rd^3r^\prime}\\
B_3&=\frac{15}{32\pi^2r_0^5A^{5/3}}\int\int_V\frac{d^3rd^3r^\prime}{\left|\vec{r}-\vec{r}^\prime\right|}
\left[{1-\left(1+\frac{1}{2}\frac{\left|\vec{r}-\vec{r}^\prime\right|}{a_{den}}\right)
e^{\left|\vec{r}-\vec{r}^\prime\right|/a_{den}}}\right]
\end{array}
\end{equation}
These shape factors, along with their derivatives with respect to the radius are solved numerically in the FRDM,
and the balance between the Coulomb energy, compressibility, and the surface tensions are found based on
shape parameters taken from the deformation parameters in a folded Yukawa potential.  For a 
spherical geometry, $B_1$ and $B_3$ are equal to 1 with the ranges $a$ and $a_{den}$ are equal to 0 - a sharp 
surface.
The macroscopic constants in the FRDM have been determined based on a non-linear least squares fit requiring
about 1000 iterations of the process of choosing parameters, recalculating potential surfaces, finding
the fit, and re-adjusting parameters. 

The microscopic portion of the model involves the individual shell terms plus the effective interaction pairing
gaps for protons and neutrons\cite{moller92,pradhan73}:
\begin{equation}
\label{pairing}
\begin{array}{l@{\quad}l}
\Delta_{G_n} & = \frac{r_{mic}B_s}{N^{1/3}}\\
\Delta_{G_p} & = \frac{r_{mic}B_s}{Z^{1/3}}\\
\end{array}
\end{equation}
in which a constant level density is assumined in the vicinity of the Fermi energy level.  The value of
$r_{mic}$ is determined by a least-squares fit. 

\begin{figure}
\includegraphics[angle=90,width=15cm]{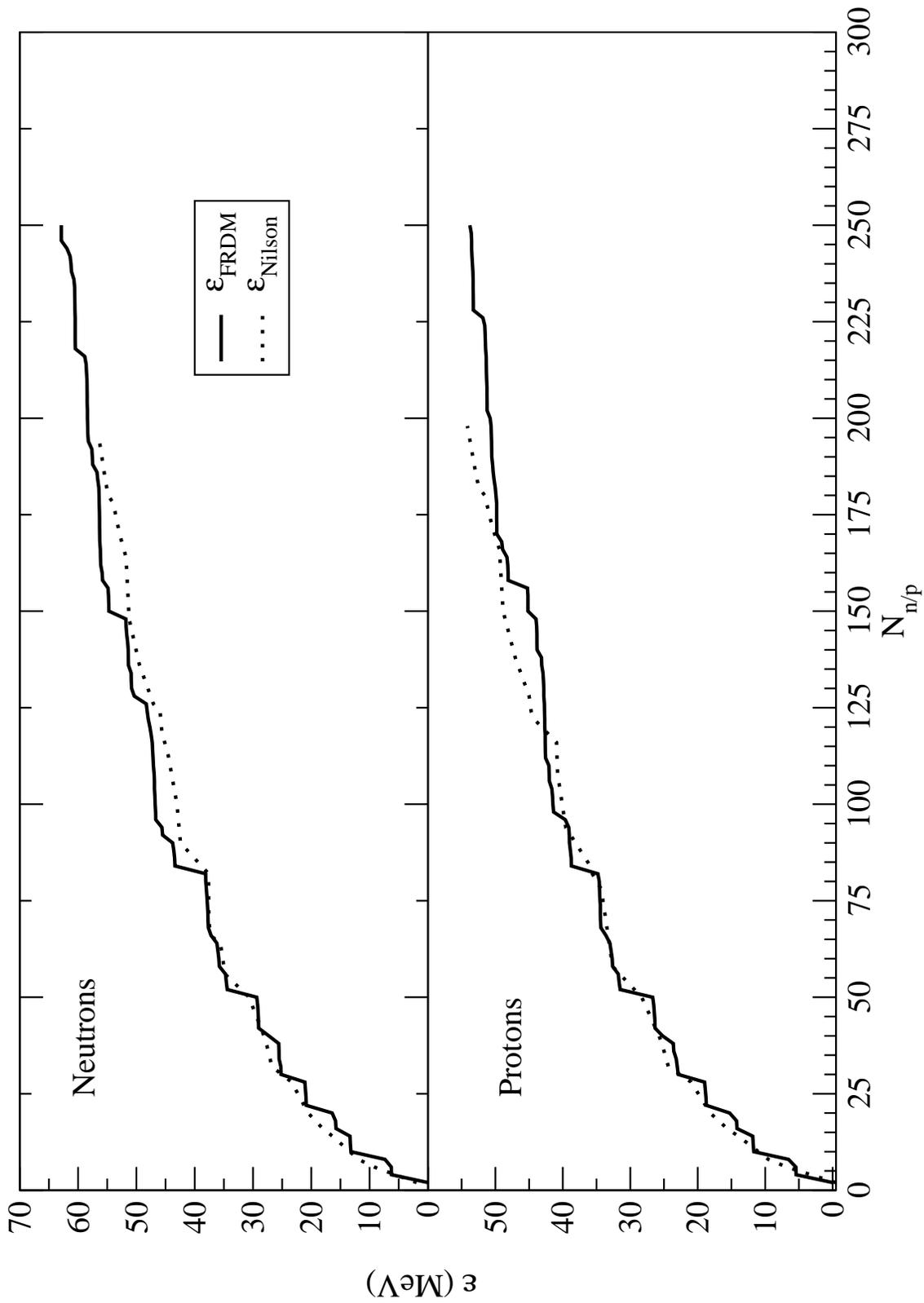}
\caption{\label{shells} Example of two-particle levels for $^{132}$Sn neutrons (top graph) and
protons (bottom graph) 
in the FRDM with LN pairing (solid line).  The dotted line is that of the Nilsson model with no 
deformation.}
\end{figure}

Finally, microscopic corrections to the FRDM are carried out by decoupling the shell and pairing potentials
as well as the proton and neutron potentials.  The single-particle potential is then the sum of
the central potential $V_1$, which - as stated - is the folded Yukawa potential, the spin-obit potential
$V_{s.o.}$, and
the Coulomb central potential for protons $V_C$.  Two-particle levels as a function of nucleon number are compared to 
those determined by the Nilson model (with no deformation) for the $^{132}$Sn nucleus in Figure \ref{shells} showing
a reasonable agreement between the two.

For a more complete description of the LN pairing 
interaction and microscopic interaction, the reader is referred to references \cite{moller00} and to 
\cite{pradhan73} and references therein.

In the present calculation, single particle levels are determined for both the parent and 
daughter nucleons.  That is, transitions between specific levels are calculated for a 
nucleus with j parent nucleons and k daughter nucleons to a nucleus with j-1 parent nucleons 
and k+1 daughter nucleons.  Thus, corrections to any shell and pairing energies are 
intrinsically accounted for.  
\subsubsection{Quantum Numbers of Single Particle States}
\label{qnum}
\begin{figure}
\includegraphics[width=16cm]{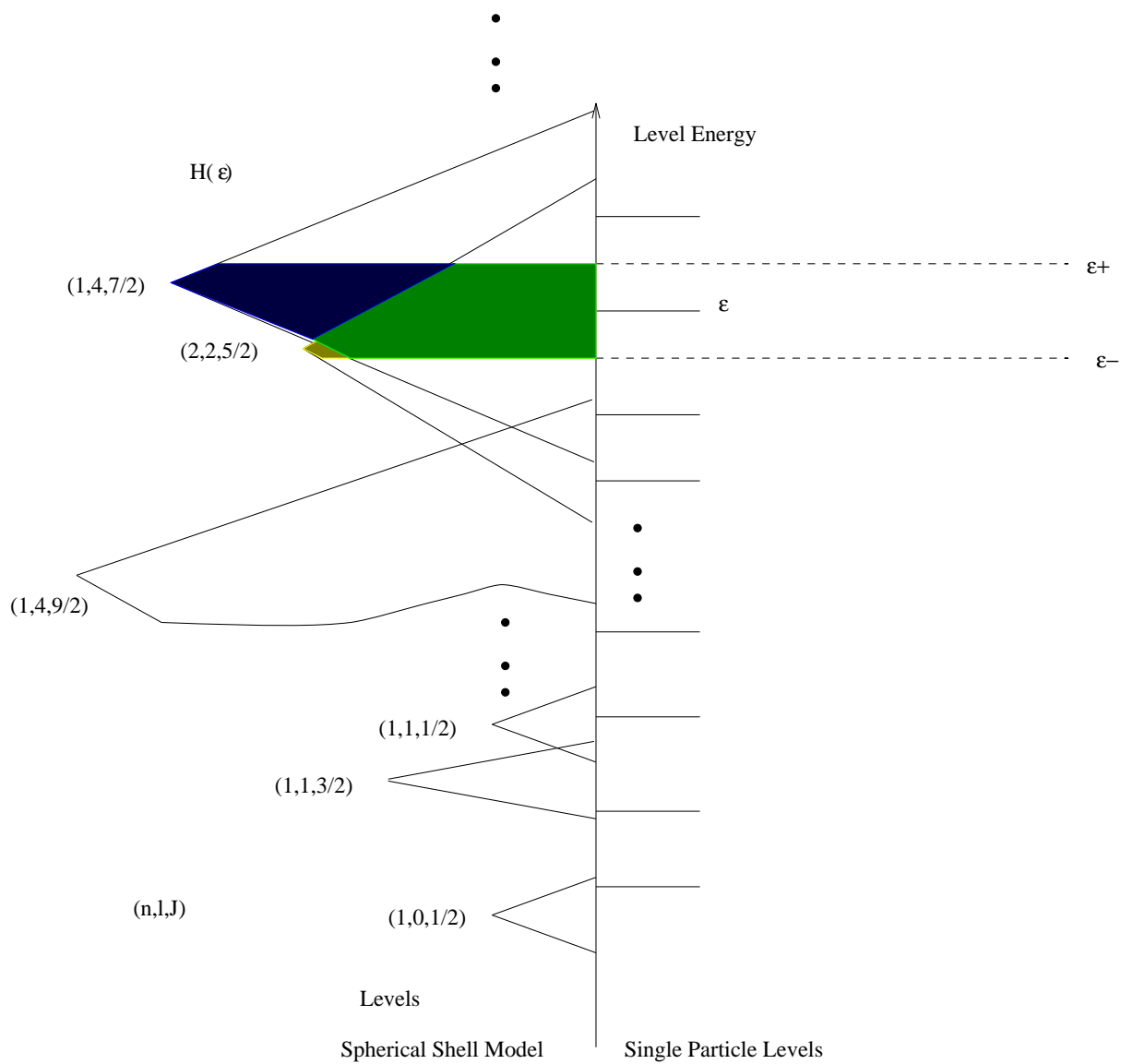}
\caption{\label{eigen} Sample assignment of quantum numbers based on level 
proximity 
to 
levels of the spherical shell model \cite{nakata97}.  For level
$\varepsilon$, a contribution from two spherical shell model levels
is made, and the integration is over the shaded regions in each
triangular region corresponding to the level.  Note that the level
height corresponds to the level degeneracy, and the widths scale with
standard level spacing.
}
\end{figure}

A knowledge of the quantum numbers of each level enables one to calculate the
order of a particular $\beta$ transition.  However, as stated previously, a 
Fermi Gas model level at a specific energy may correspond to several shell
model levels in the present formulation, so it may turn out to be a mixture
of single-particle shell model eigenstates.
The basis of states is assumed to be those of a spherical shell model with a
Woods-Saxon potential \cite{shirley96}.  The weight for each set of 
quantum numbers in a level
is determined by the proximity of a level's energy to that of a basis state.
The overlap of the eigenfunctions at a shell level determines the strength of
their presence in the level (Figure \ref{eigen}).  The method of Nakata et al.
is used to determine not only the quantum numbers
of parent nucleons, but also those of daughter nucleons \cite{nakata97}.  
Triangular distribution functions centered on the energy of the 
states are defined:
\begin{equation}
\label{eq21}
H_{n,l,j}(\varepsilon)=\left\{
\begin{array}{l@{\qquad \qquad}l}
\frac{N_{n,l,j}}{q(A)}\left[1-\frac{|\varepsilon-\varepsilon_{n,l,j}|}
{q(A)}\right] & -q(A)<\varepsilon-\varepsilon_{n,l,j}<q(A)\\~\\
0 & \mbox{otherwise}
\end{array}
\right.
\end{equation}
where $\varepsilon_{n,l,j}$ is the energy of the eigenstate corresponding to
the usual quantum numbers.  The function is already normalized to N$_{n,l,j}$,
which is equal to the degeneracy of a level, 2j+1.  The width q(A) of the
basis function of the $i^{th}$ level in a nucleus with $\mu$ levels is given by:
\begin{equation}
\label{eq22}
q(A)=6\frac{d_i}{d_\mu}
\end{equation}
where d$_i$ is the standard level spacing:
\begin{equation}
\label{eq_di}
d_i\equiv\frac{2}{3\left({2i-2}\right)}\varepsilon_i^0
\end{equation}
where $\varepsilon_i^0$ is the energy of the Fermi surface of the $i^{th}$ level.  Since 
the values of $\varepsilon$ are allowed to vary smoothly, and scale smoothly with level, 
Fermi gas levels are appropriate to effect this scaling.

The total sum of the functions in Equation \ref{eq21} is also defined:
\begin{equation}
\label{eq23}
G\left(\varepsilon\right)=\sum_{n,l,j}H_{n,l,j}(\varepsilon)
\end{equation}
And G($\varepsilon$) is automatically normalized to the total number of
nucleons.
To find the coefficient $\omega$(i;n,l,J) in Equation \ref{eq9} for the 
i$^{th}$ single-particle level, it is necessary to normalize the function
G($\varepsilon$)
such that the overlap in the the region of the level is equal to two - the 
number of nucleons per level:
\begin{equation}
\label{eq24}
N_i\int_{\varepsilon^-}^{\varepsilon^+}G(\varepsilon) \, 
d{\varepsilon}=2
\end{equation}
where the integration limits are the midpoints between the i$^{th}$ level and
adjacent levels:
\begin{equation}
\label{eq25}
\varepsilon^{\pm}=\frac{\varepsilon_i+\varepsilon_{i\pm1}}{2}
\end{equation}
For i=1 $\varepsilon^-$=0.  The coefficient for a particular configuration
of quantum numbers in the i$^{th}$ level is then:
\begin{equation}
\label{eq26}
\omega(i;n,l,j)=\frac{1}{2N_i}\int_{\varepsilon^-}^{\varepsilon^+}
H_{n,l,j}(\varepsilon)\, d\varepsilon
\end{equation}
This method is diagrammed in Figure \ref{eigen} \cite{nakata97}. 
The degeneracy of the spherical shell model level $\varepsilon_{n,l,j}$ is
already accounted for in Equation \ref{eq21} with the factor N$_{n,l,j}$.
The selection rules in Equation \ref{eq10} are then given by the 
function $\Lambda_\Omega(\zeta;\xi)$ and can now be easily determined, where
$\Omega$ is the order and rank of the transition ($\Delta{l}$,$\Delta$j); 
$\zeta$ and $\xi$ represent the quantum numbers of the initial and final 
nucleon states respectively, (n$_i$,$l_i$,j$_i$;n$_f$,$l_f$,j$_f$).  Thus, 
the selection rules can be simply stated as:
\begin{equation}
\label{eq26a}
\Lambda(n_i,l_i,j_i;n_f,l_f,j_f)=
\delta\left({l_f,l_i{\pm}\Delta{l}}\right)
\delta\left({j_f,j_i\pm\Delta{j}}\right)
\end{equation}
where the values of $\Delta{l}$ and $\Delta{J}$ depend on the initial and final
state quantum numbers.

Because each state's wave function is a sum of shell-model eigenstates, the
smearing of single-state quantum numbers over several single-particle states is
possible.  The net result is still an average calculation of 
$\beta$-decay rates, but
now individual particle and particle-hole excitations can be simulated by promoting an
arbitrary nucleon to any available level.  Further, the selection rules are now easily 
calculated using the method of the previous section.
\subsubsection{Corrections in Pairing Energy Due to Decay}
\label{correction}
If the highest-level paired neutron decays, the other neutron in that level
gains an amount of energy roughly equal to the pairing energy of that level.
As well, paired neutrons in the next lowest level are 
then expected to
reconfigure by both losing an amount of energy equal to the pairing energy of
that level.  Thus, the endpoint energy (taken to be positive in this
respect) is changed by an amount equal to 2b$_{\mu-1}$-b$_\mu$ just from the
changes in the parent nucleus.  (It is not
necessary to include the energy of the decaying neutron since this is 
accounted for in the quantity $\varepsilon_i-\varepsilon_f$.)

Similarly, in the daughter nucleus, if the highest proton level 
(in $\beta^-$-decay) is unpaired, and
a neutron decays to that level, then the highest original paired level 
loses its
pairing energy for the two protons, while the level to which the neutron
decays becomes the highest paired level, increasing the total endpoint
energy by an amount b$_\mu$, and the total energy available to the ejected
electron changes by b$_\mu$-2b$_{\mu-1}$.  (Again, the pairing energy of the
decaying particle is not included for the same reasons mentioned 
above.)
\subsection{Results of $\beta$-Decay Rate Calculations}
\label{results}
\begin{figure}
\includegraphics[angle=90,width=15cm]{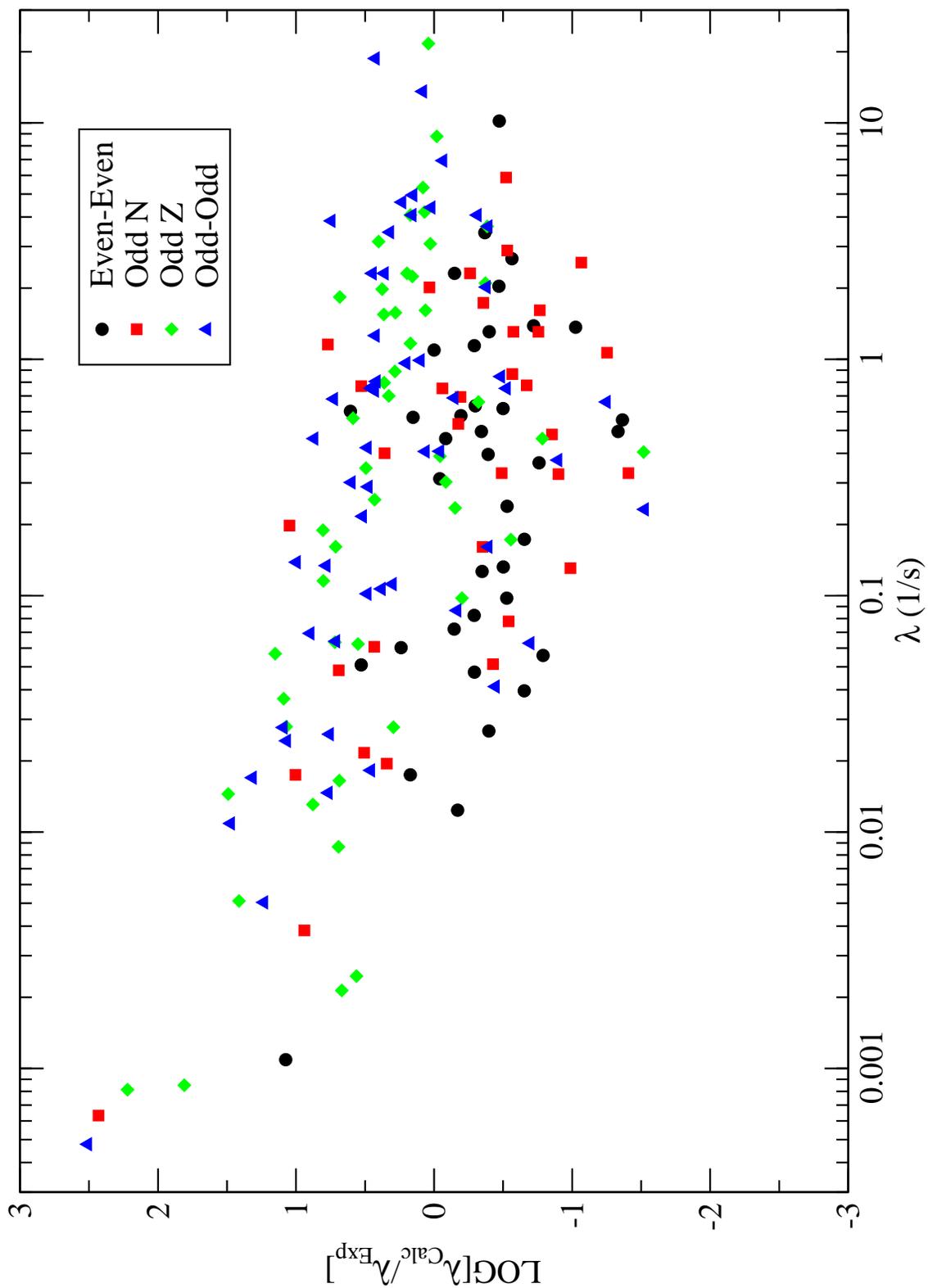}
\caption{\label{FRDM_rates} Error in calculated $\beta$-decay rates as a 
function of the known half-life using the FRDM \cite{beta_FRDM2002}.}
\end{figure}

\begin{figure}
\includegraphics[angle=90,width=15cm]{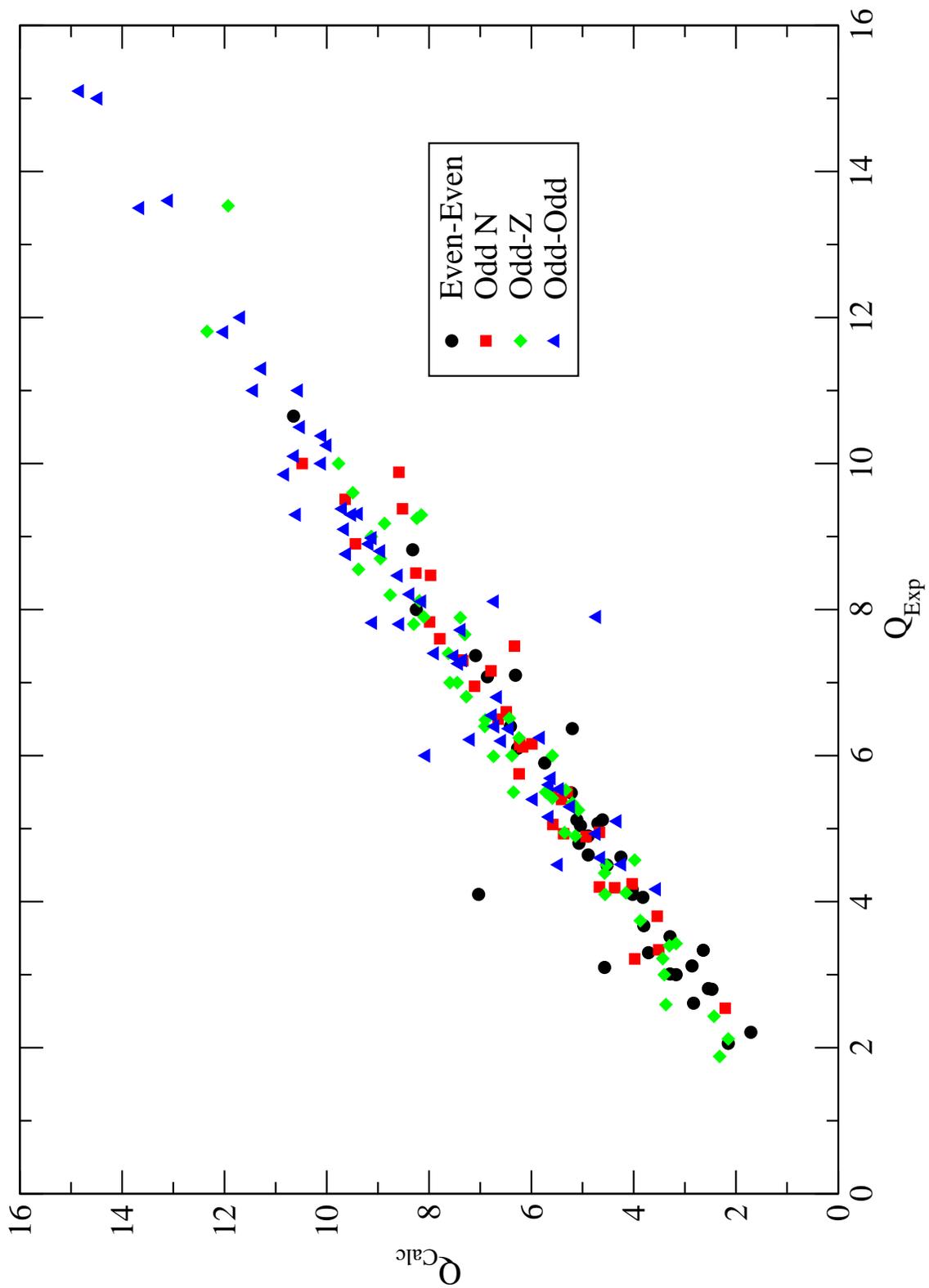}
\caption{\label{FRDM_q} $\beta$-Decay Q-values calculated with the FRDM compared to 
experimentally known Q-values.}
\end{figure}

Decay rates and Q-Values have been calculated with the FRDM in previous works
\cite{beta_FRDM2002}.   The level of accuracy of these calculations is shown in Figures 
\ref{FRDM_rates} and \ref{FRDM_q} for about 200 sample neutron-rich nuclei unstable against $\beta^-$decay for 
which the decay rates and Q-values are experimentally known.  In Figure \ref{FRDM_rates}, 
the value of S is defined as:
\begin{equation}
\label{eq1-4}
S{\equiv}\log{\left(\frac{\lambda_{\mbox{calc}}}{\lambda_{\mbox{exp}}}\right)}
\end{equation}  
In each figure, most calculated rates fall within an order of magnitude of the
true rate.
\subsubsection{Excited State Decays}
\label{excited}
A next logical step is to calculate decay rates for nuclei in excited states.
The present formulation gives no preferential treatment to nucleons in
single-particle excited states.  One simply needs to know the values of the 
excited levels and the vacant levels for Equations \ref{eq7} and \ref{eq8}.  The
levels are already calculated from \S \ref{spcalc}.  Levels can be 
calculated beyond the maximum filled level in the ground state, although the
simplicity of the model can render these values slightly  different from their
experimental counterparts.  

In stellar environments, the probability of finding a nucleus in an 
excited state at a given temperature can be obtained from the partition 
function by knowing the spin and energy of the excited state.  Since each 
particle level holds only two nucleons, the degeneracy of each level is two,
and the quantum numbers are given
by the level's proximity to those of the spherical shell model (as 
described in \S\S \ref{qnum}).
The difference in energy:
\begin{equation}
\label{eq28}
{\Delta}E={\sum_{i_{N,Z}=1}^mn_{n,i}^*\varepsilon_i^* - 
\sum_{i_{N,Z}=1}^{\mu}n_{n,i}^0
\varepsilon_i^0}
\end{equation}
where values with an $^*$ are those for an excited-state nucleus, while those
with a $^0$ are those corresponding to a ground-state nucleus.  The summation 
is taken over neutron and proton levels.  The value m is
the
highest bound state level.  The first term represents the total energy of the 
excited nucleus, and the second term is the total energy of the ground-state
nucleus.  The average decay rate of an isotope in a
stellar environment is then the weighted sum over energy states \cite
{takahashi78}:
\begin{equation}
\label{eq29}
\lambda = \sum_{{\Delta}E}P({\Delta}E)\lambda({\Delta}E)
\end{equation}
where P($\Delta$E) is calculated using the partition function.

\begin{figure}
\includegraphics[angle=90,width=15cm]{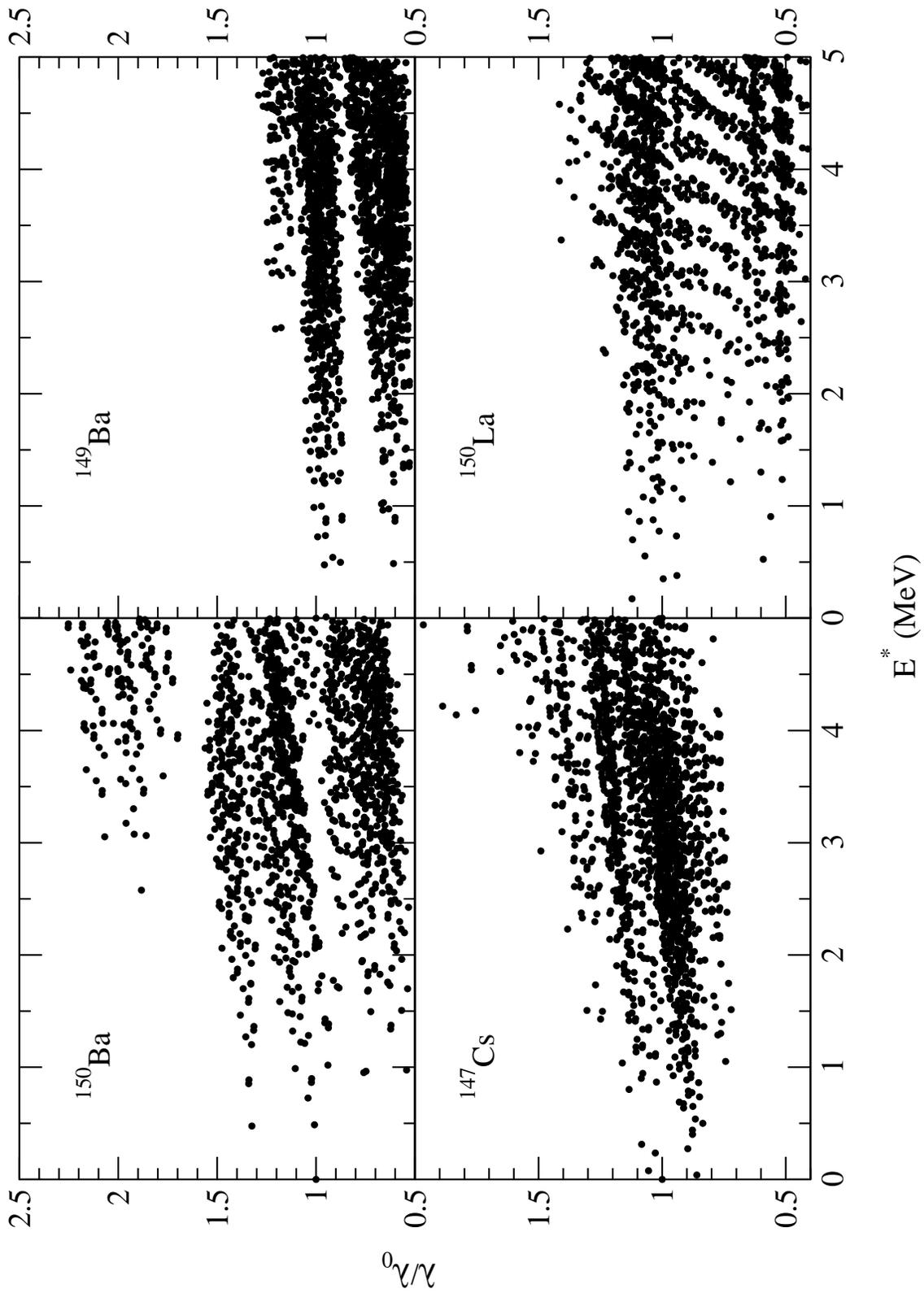}
\caption{\label{beta_states} $\beta$-Decay rates of four nuclei as a function of excited state 
energy of individual single-particle excitations.}
\end{figure}

Excited states and their decay rates have been calculated for several nuclei.  A sample of 
the decay rates as a function of these states are shown for four nuclei in Figure 
\ref{beta_states}.  This figure shows the dependence of decay rates on excitation for an 
even-even nucleus ($^{150}$Ba), an odd-N nucleus ($^{149}$Ba), an odd-Z nucleus ($^{147}$Cs),
and an odd-odd nucleus ($^{150}$La).  One first notes the multiplicity of levels resulting 
from the splitting of states with spins higher than 1/2, a direct result of 
using levels with a degeneracy of two.

One also notes the apparent clustering of decay rates into bands.  In the case of 
$^{150}$Ba, one notes three major bands at $\frac{\lambda}{\lambda_0}\approx$0.75, 1, and 
2 where $\lambda_0$ is the ground-state decay rate.  
This is due to the fact that single-particle levels are dominated by specific 
shell-model eigenstates in certain regions, as expected.  In the case of $^{150}$Ba, the first several 
single-particle neutron excitations above the ground-state contain large admixtures of the 
$i_{13/2}$ state, while those near the ground-state Fermi surface are predominantly in the 
$h_{9/2}$ state.  In the case of transitions in this region, the decay order of the 
particle-hole configurations goes from GT to first-forbidden, thus reducing the decay rate.  
Single-particle excitations in 
certain energy regions are expected to have similar decay strengths.  The band at 
$\frac{\lambda}{\lambda_0}\approx$2 is due to proton $\bar{\pi}g_{9/2}\pi{g}_{7/2}$ 
particle-hole configurations which do not change the transition order, but change the 
decay Q-value by opening up a hole at lower energies.  
The band at $\lambda/\lambda_0\approx$1 corresponds to $\bar{\nu}i_{13/2}\nu{i}_{11/2}$ configurations.
Since this has very little impact on the highest-energy GT (or first-forbidden) transitions 
(to the $\pi{g}_{7/2}$ or $\pi{g}_{9/2}$ state), the overall transition rate is changed very
little.  It is expected that higher excitations will result in other band 
structures.  However, since astrophysically interesting temperatures are below 1 MeV, the contribution to
average decay rates from states much above this is small, so only lower excitations were
studied.  

Similarly, the band 
structures in the other nuclei shown can be explained in a similar manner.  The band structures 
of other nuclei vary based on their individual structures.  For example, $^{150}La$ has a band 
at $\frac{\lambda}{\lambda_0}<1$, but no band at $\frac{\lambda}{\lambda_0}>1$ (for the 
energies of interest).  This nucleus has a higher deformation than $^{150}$Ba, as well as 
an unpaired neutron and an unpaired proton.  The lowest excitation energies are accomplished 
via neutron $\bar{\nu}h_{9/2}\nu{i}_{13/2}$ particle-hole configurations, but single-proton
excitations are due primarily to the unpaired proton, which has little bearing on the
available hole, but does account slightly for the vertical band structure in the figure; the rate increases 
with the transition energy for a single type of transition.

\begin{figure}
\includegraphics[angle=90,width=15cm]{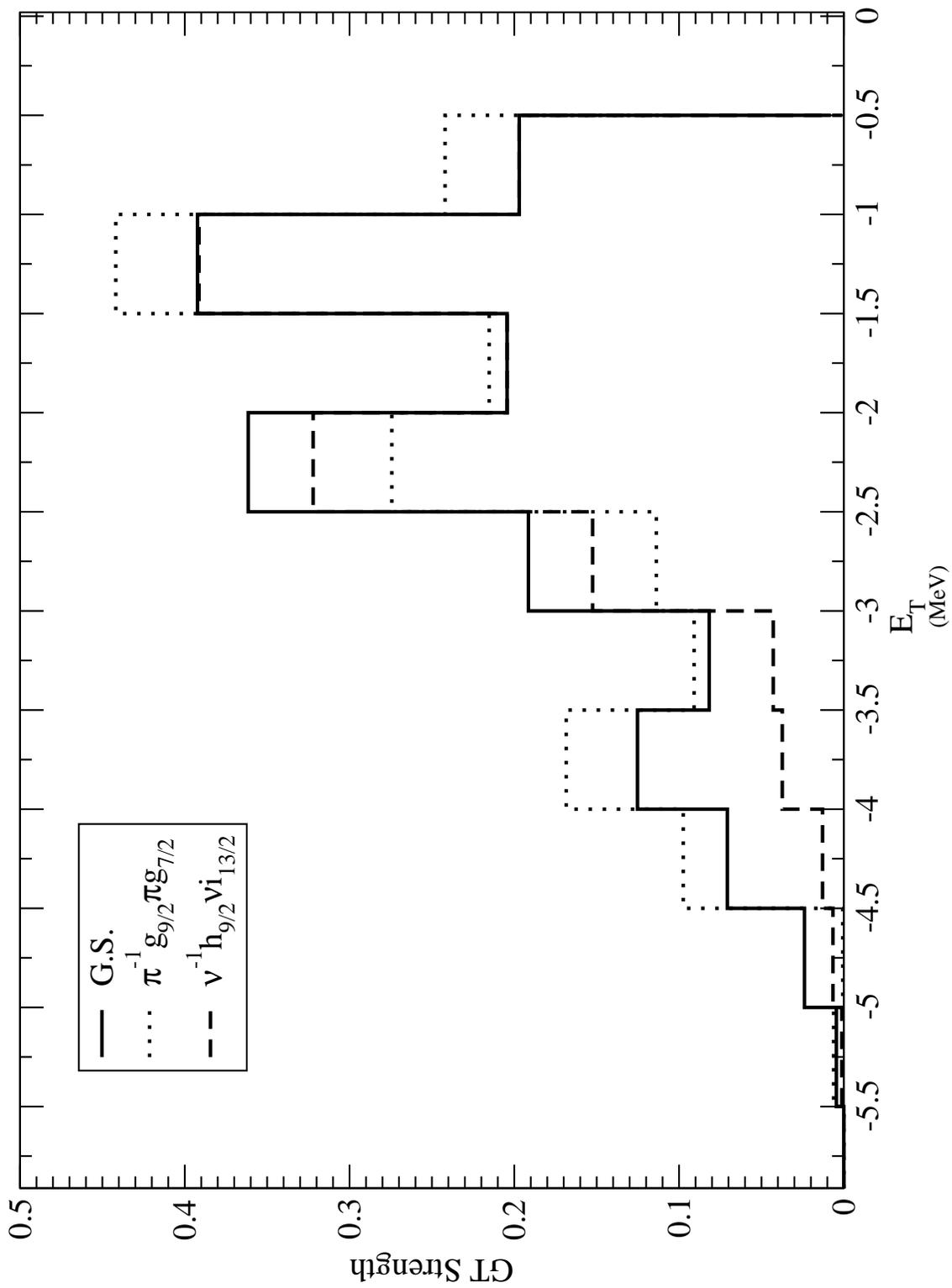}
\caption{\label{exc_func} GT transition strength function for various particle-hole states in 
$^{150}$Ba.}
\end{figure}

\begin{figure}
\includegraphics[angle=90,width=15cm]{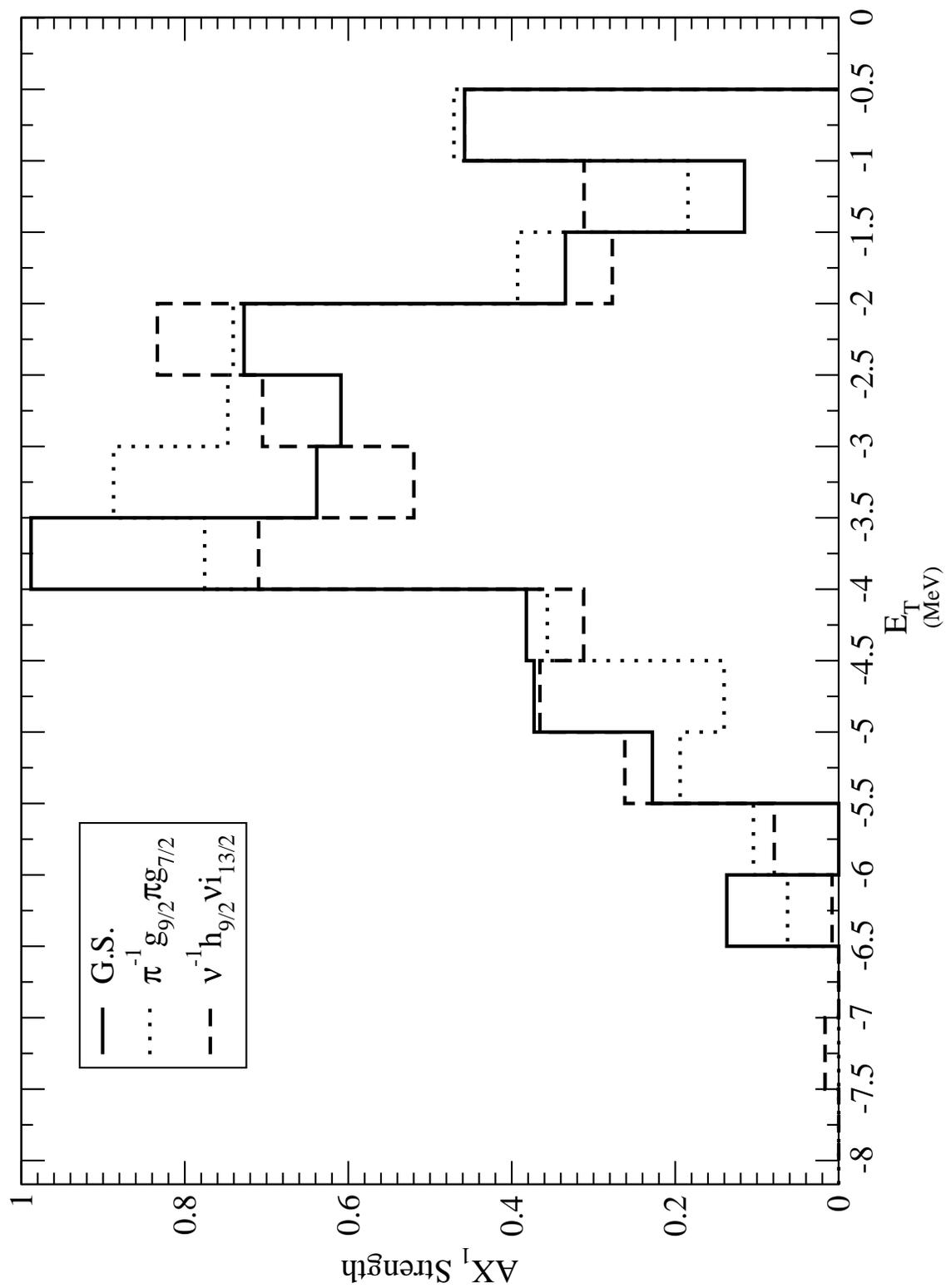}
\caption{\label{exc_func_ff} First-forbidden transition strength function for various 
particle-hole states in $^{150}$Ba.}
\end{figure}

Of course, excitations can change the strength functions, as the order of a particular 
transition can be altered.  In the case of $^{150}$Ba, the GT strength function is shown for 
the ground state, a transition to the region dominated by the $\bar{\pi}g_{9/2}\pi{g}_{7/2}$ 
state, and an excited state dominated by the $\bar{\nu}h_{9/2}\nu{i}_{13/2}$ state in Figure 
\ref{exc_func}.  As expected, the neutron excitation lowers the overall GT transition strength, 
particularly at the highest transition energies, where the transition rate is highest.  
Conversely, the proton excitation increases the overall rate slightly at the lowest energies, 
while also increasing the Q-value.  It should be noted in this case that, although the rate is 
doubled and the GT strength function for the proton excitation looks very similar to that of 
the ground-state, the average lifetime is still relatively low ($\sim$300ms), and the result is 
a small absolute increase in lifetime (to about 315ms).  
Overall, there are similar small changes in the first-forbidden strength functions, shown in 
Figure \ref{exc_func_ff}.

\begin{figure}
\includegraphics[angle=90,width=14cm]{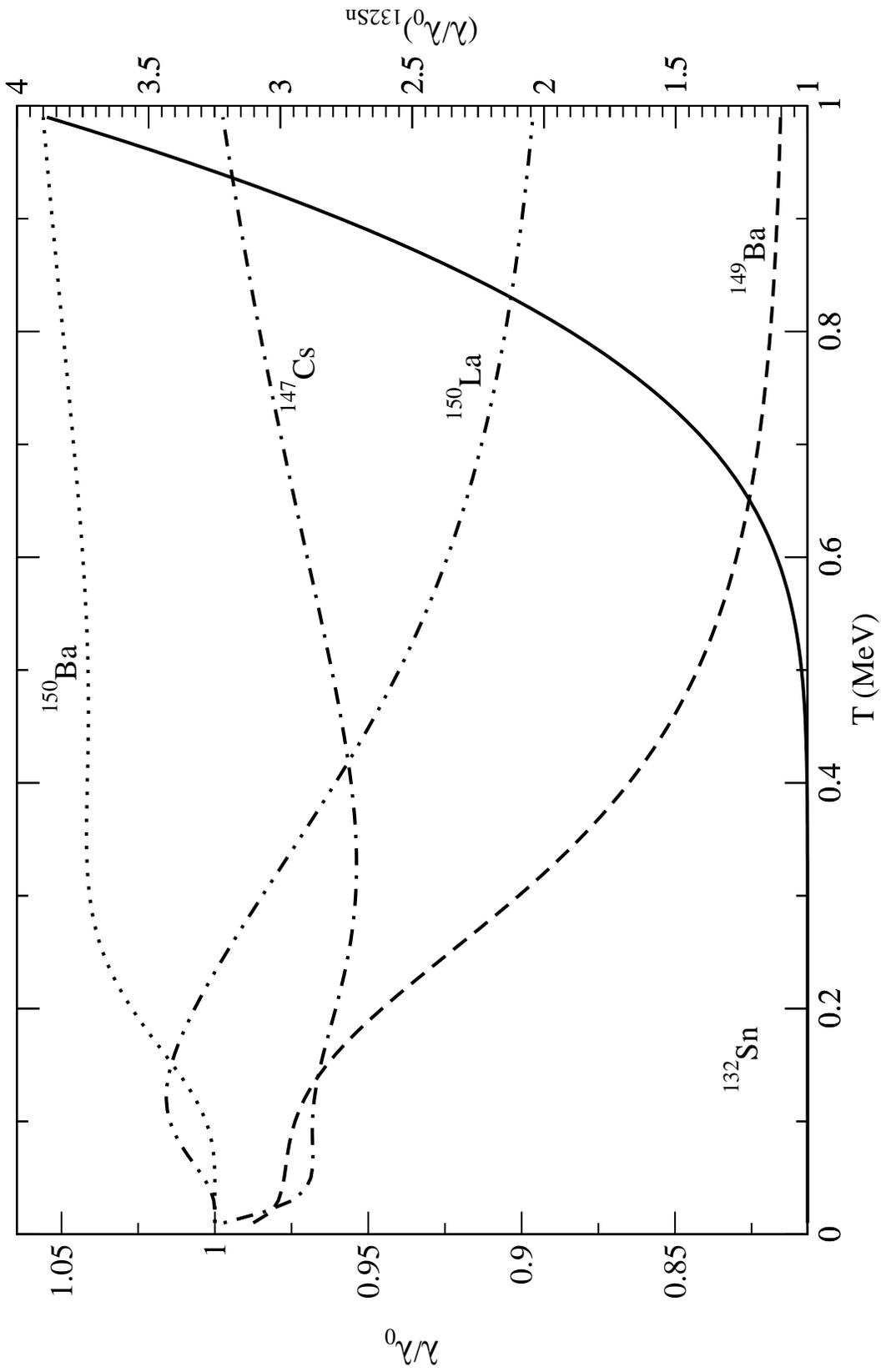}
\caption{\label{beta_temp} $\beta$-Decay rates of the nuclei in Figure \ref{beta_states} as a 
function of temperature.  The decay rate of $^{132}$Sn is also plotted as a function of 
temperature.}
\end{figure}

Given these figures, one cannot definitively state that excited-state nuclei decay at a higher 
rate than ground-state nuclei, as the transition order shift becomes important.  The average decay rates 
of the nuclei in Figure \ref{beta_states} are plotted as a function of temperature in Figure 
\ref{beta_temp}, calculated using Equation \ref{eq29}.  The decay rate of $^{132}$Sn as a 
function of temperature is also plotted.  While each nucleus behaves differently, it seems 
clear that the doubly-magic $^{132}$Sn does exhibit a sharp increase in decay rate with temperature
at higher temperature, 
most likely due to the fact that $^{132}$Sn has closed proton and neutron shells.  All 
single-particle excitations in $^{132}$Sn are quite high in energy and  
open transitions that were previously Pauli-blocked.  
The transition strength functions for the first calculated excited state of $^{132}$Sn are 
shown along with the ground-state functions in Figures \ref{gtstr}-\ref{vecstr} indicating 
greatly increased GT strengths even at high transition energy.

\begin{figure}
\includegraphics[angle=90,width=15cm]{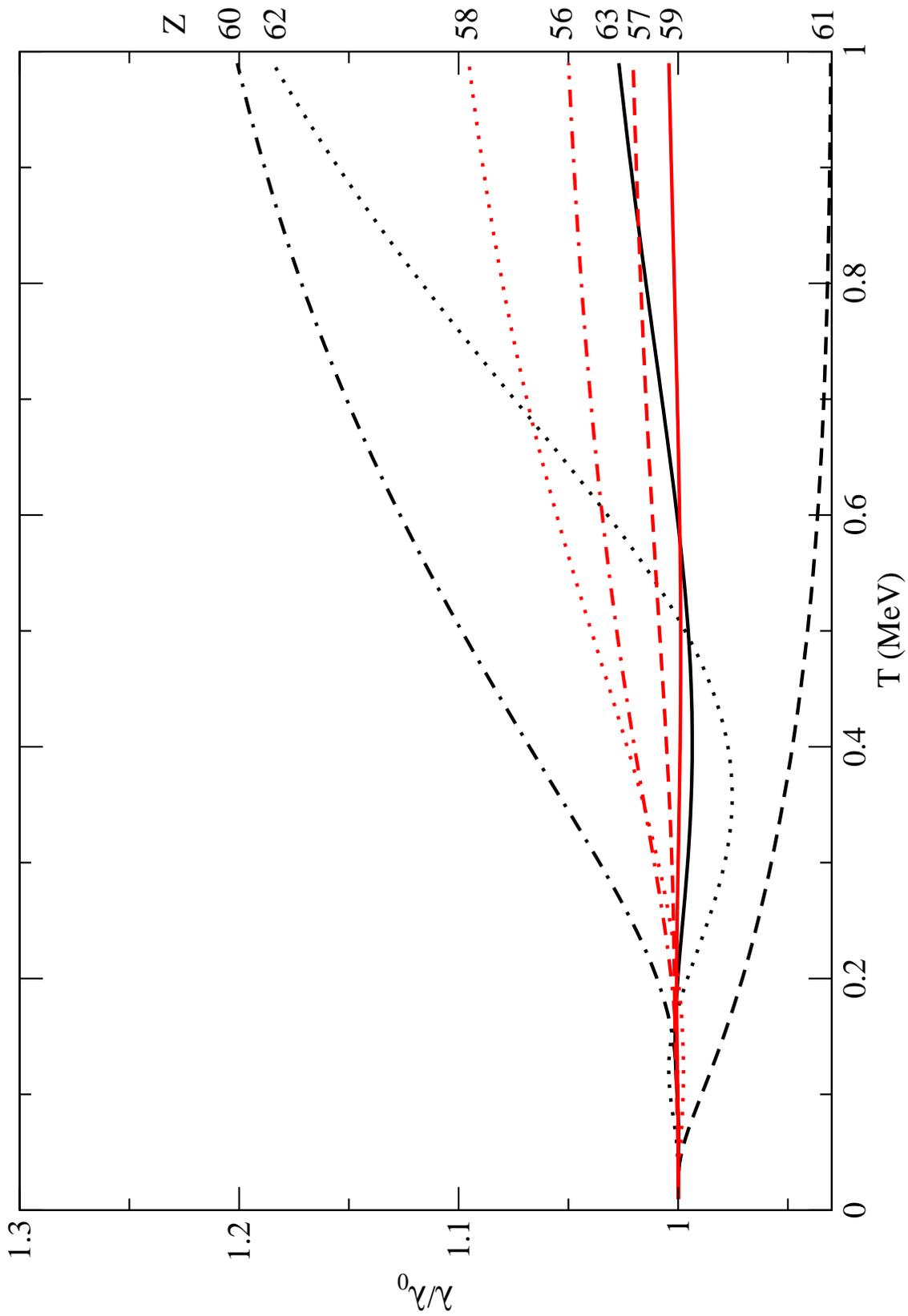}
\caption{\label{lambda_isobar}Relative decay rates for various nuclei along the A=162 isobar as 
a 
function of temperature.  In general, the nuclei closer to stability will undergo a larger
increase in decay rate with temperature.}
\end{figure}

Finally, the effect of temperature on $\beta$-decay rates is evaluated as a function of the
neutron-richness of the nucleus in Figure \ref{lambda_isobar}.  In this figure, relative
decay rates are plotted as a function of temperature for various Z on the same isobar A=162.
For this isobaric chain, there may be  a slight dependence on the stability of the nucleus.  
While more study is warranted, this may be of interest in suggesting that 
GT
transitions of the highest level neutrons in the
very neutron-rich nuclei along the r-process path are most likely not Pauli blocked, as they
might be for the nuclei closer to stability.  In this case, single-proton excitations will
occupy states that a neutron would otherwise decay to.  Nuclei closer to stability, on 
the other hand, have truncated GT strength functions.  Thus, proton or neutron excitations
are more likely to open transitions that were originally Pauli-blocked.
\section{Effects of Excited-State Decays on the r-Process}
\label{r-process}
Relative changes in decay rates along the r-process path are shown in Figure \ref{rel_rates} for
the line of constant neutron separation energy $S_{n}\approx2.5$ MeV  for three different
temperatures.  The effect is to increase slightly the relative ratios of the ground-state rates 
at lower mass.  However, this effect is only slight ($<$10\%)
in temperature regimes relevant to the r-process.  A network calculation is
necessary to evaluate fully the magnitude of the effect on the r-process.  The increase
in the $^{132}$Sn rate may be enough to make a significant difference in the final abundances,
but this is not clear yet.
\subsection{The Explosion Model}
The r-process environment used in the present study is constructed in
the semi-analytic model of the nutrino-driven winds of reference \cite{otsuki00}.
The
mass flow out of the supernova hot-bubble region is treated as
a spherically symmetric, steady flow around a neutron star 
of mass M and
radius R in the Schwarzchild geometry, which
is described \cite{shapiro83} by the following sets of equations:
\begin{eqnarray}
\label{greqns}
\dot{M} & = & 4\pi{r^2}\rho{u}
\nonumber\\
u\frac{du}{dr} & = & \frac{1}{\rho_T+P}\frac{dP}{dr}\left(1+u^2-\frac{2M}{r}
\right)-\frac{M}{r^2}
\nonumber\\
\dot{q} & = & u\left(\frac{d\varepsilon}{dr}-\frac{P}{\rho_b}\frac{d\rho_b}
{dr}
\right)
\end{eqnarray}
where $\dot{M}$ is the mass outflow rate, $\rho_b$ is the baryon density,
$\varepsilon$ is the specific internal energy, $u$ is the radial component of 
the
four-velocity, and $\dot{q}$ is the net heating rate; $\dot{q}$ is positive
just above the surface of the neutron star, decreasing roughly 
exponentially 
with radius.  The total energy density 
$\rho_T$ is related to the baryon density by $\rho_T=
\rho_b\left({1+\varepsilon}\right)$.
In the above equations, the constants $\hbar$, $c$, $k_B$, and $G$ are equal 
to
unity.  In a stationary rest frame, the radial velocity of the expanding 
bubble is
the product of the Lorentz factor and the radial component of the 
four-velocity:
\begin{equation}
v_r=u\left(1+u^2-\frac{2M}{r}\right)^{-\frac{1}{2}}
\end{equation}

The net heating rate is determined by the sum of neutrino heating and cooling 
from the following reactions and their inverses:
\begin{eqnarray}
\label{heating}
\nu_e+n & \leftrightarrow & e^- + p
\nonumber\\
\bar{\nu}_e + p & \leftrightarrow & e^+ + n
\nonumber\\
\nu_i + e^- & \leftrightarrow & \nu_i + e^-
\nonumber\\
\nu_i + e^+ & \leftrightarrow & \nu_i + e^+
\nonumber\\
\bar{\nu}_i + e^- & \leftrightarrow & \bar{\nu}_i + e^-
\nonumber\\
\bar{\nu}_i + e^+ & \leftrightarrow & \bar{\nu}_i + e^+
\nonumber\\
\nu_i + \bar{\nu}_i & \leftrightarrow & e^- + e^+
\end{eqnarray}
where the index $i$ corresponds to each neutrino flavor.  The Newtonian
forms of these 
heating 
and cooling rates are described in detail in references \cite{woosley94} and \cite{qian96}.
Since the mass flow is influenced by the properties of the
proto-neutron star, the heating rate is strongly dependent on the neutrino
luminosities and energies.  Transformation of the solid angle from the
reference frame of the neutron star to the r-process site leads to the 
correct form of the heating rate in the Schwarzchild geometry
\cite{otsuki00}.  In this calculation, the 
neutrino luminosity 
$L_{\nu_i}$ is
assumed to be the same for all neutrino flavors and the RMS energies are used;
the energies for $\varepsilon_{\nu_e}$, $\varepsilon_{\bar{\nu}_e}$,
$\varepsilon_{\nu_{\mu,\tau}}$, and $\varepsilon_{\bar{\nu}_{\mu,\tau}}$ are
12, 22, 34, and 34 MeV respectively.

Equations \ref{greqns} are supplemented with the equations of state in the
high-temperature limit taken from Qian and Woosley (1996):
\begin{eqnarray}
\label{eqofstate}
P & = & \frac{11\pi^2}{180}T^4 + \frac{\rho_b}{m_N}T
\nonumber\\
\varepsilon & = & \frac{11\pi^2}{60}T^4 + \frac{3}{2}\frac{T}{m_N}
\end{eqnarray}
where $m_N$ is the nucleon rest mass.

Using Equations \ref{greqns} and \ref{eqofstate}, the evolution of the hot
bubble is followed.  The adjustable parameters for solving the equations are
$M$, $R$, and $\dot{M}$.  The neutrino luminosity is taken to be 
5 - 7$\times$10$^{51}$erg s$^{-1}$.  The 
equations are solved implicitly  and the result is the velocity and 
thermodynamic quantities as a function of r (and hence, t).  
Any trajectory (velocity, temperature, and density as a function of radius)
can be specified with
the parameter set, and the entropy and dynamic timescale - parameters 
important 
to the r-process - are direct results of the evolution of the r-process
environment.  In this particular calculation, the approximation
of a steady flow is used \cite{otsuki00}, which allows
an analysis of the rarified region about the supernova core without 
requiring a
tremendous amount of computational power.

For this model, a typical
outer boundary temperature of T=0.1 MeV behind the shock at r$\approx$10,000 km,
consistent with the theoretical result of the 
benchmark hydrodynamic simulations of the delayed-explosion of Type II
supernovae \cite{woosley94} is used.  Subsequent extended studies of the neutrino-
driven wind \cite{wanajo01,wanajo07} have used the outer boundary condition that
the mass ejection rate $\dot{M}$ (Equation \ref{greqns}) is taken to
be 99\% of its critical value (that is, the value for which wind velocity is 
supersonic).  We therefore adopt the same
outer boundary condition as that of ref. \cite{wanajo01} with the core radius
held constant at 10 km in the present calculations is used.  
The
only two remaining hydrodynamical parameters are the core mass and the 
neutrino luminosity.  From this, the dynamic timescale $\tau_{dyn}$, 
temperature, and
density at any radius can be determined.  From the temperature and density,
the entropy is determined. 

\begin{figure}
\includegraphics[width=15cm]{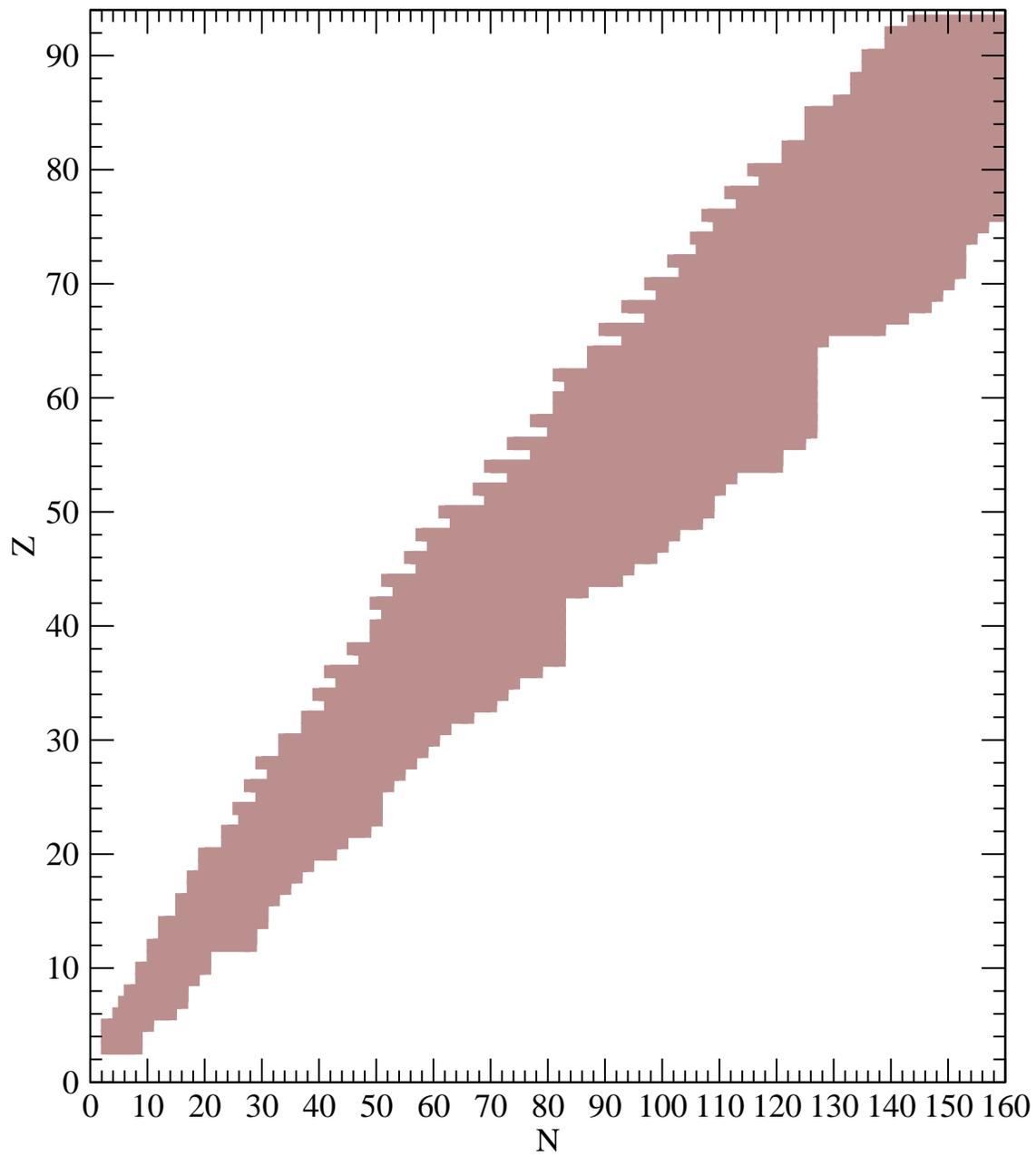}
\caption{\label{network}Nuclei included in the current nuclear reaction 
network (Terasawa et al. 2001).  Nuclei range from stable nuclei, 
(those on the left side of the
indicated region), to those along the neutron-drip line (on the right side
of the region). }
\end{figure}

\begin{figure}
\includegraphics[width=15cm]{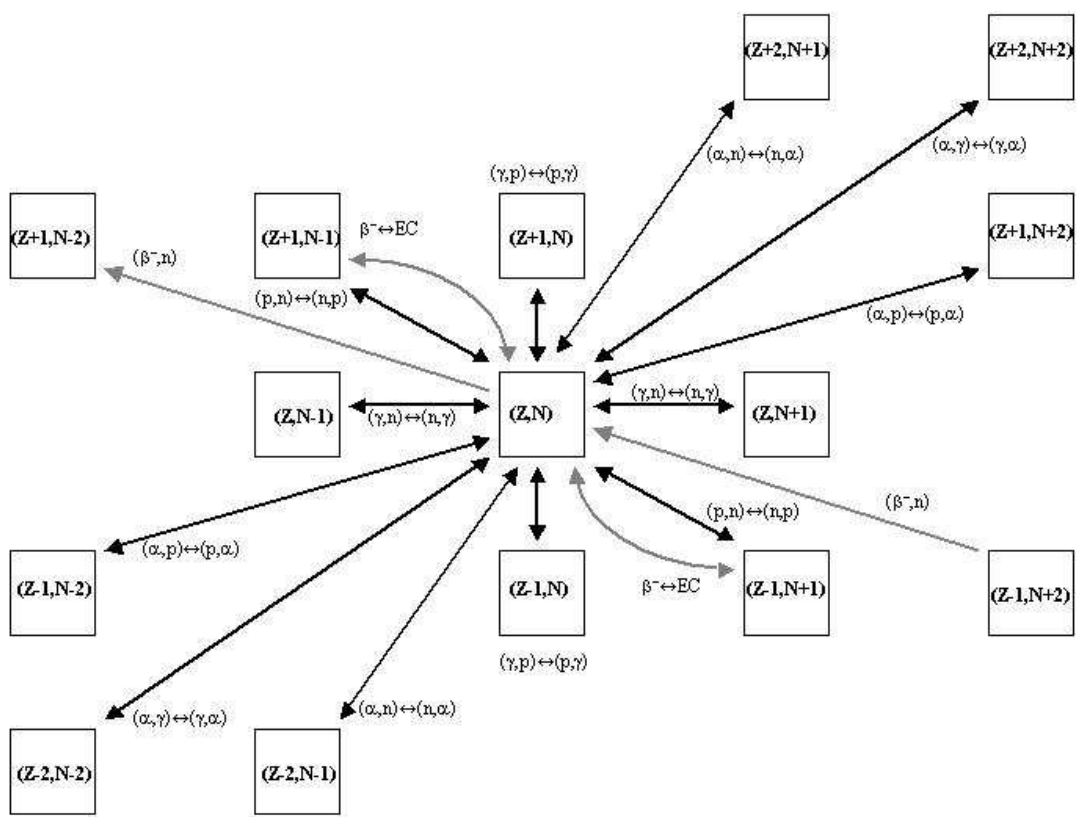}
\caption{\label{net_reacts}Possible nuclear reactions included in the
network calculation.  Each arrow corresponds to a reaction (and its
indicated inverse reaction).  Not shown are neutrino interactions.  Weak
interactions are indicated by the lighter arrows.
}
\end{figure}

The nuclear reaction network used in this calculation \cite{teresawa01,teresawa04} is coupled to the 
output of the aforementioned hydrodynamic code.  As
shown in Figure \ref{network}, the network consists of about 3700 nuclei with 
Z$\le$93 and includes all nuclei between the most proton-rich stable nuclei 
and
the neutron drip line \cite{teresawa01}.  Possible reactions in this 
network are (n,$\gamma$), 
(p,$\gamma$),
($\alpha$,$\gamma$), (p,n), ($\alpha$,p), ($\alpha$,n), $\beta$-decay,
$\beta$-delayed neutron emission, electon capture, and neutrino 
neutral- and charged-current interactions and their inverses 
\cite{teresawa01}.  These reactions are summarized in Figure
\ref{net_reacts}.  The effects of excited-state nuclei are included via
the scaling of decay rates, as discussed below.
Nucleosynthesis calculations begin at T$_9$=9, a sufficiently high 
temperature
that a state of nuclear statistical equilibrium (NSE)
exists, defined by the balance of strong and electromagnetic interactions.
At this temperature matter is mostly in the form of free nucleons.  The
initial electron fraction Y$_e$ is a parameter of the reaction network.

In order to maintain the consistency in the mass formula, the same formula 
used
to calculate shell energies for $\beta$-decay rate calculations (i.e., the 
single-particle formula of TUYY \cite{tachibana88})was used to calculate
neutron separation energies,
$\alpha$-particle separation energies, $\beta$-delayed neutron emission 
probabilities and ground-state $\beta$-decay rates
for the nuclei shown in 
Figure \ref{network}.  $\beta$-decay rates as a function of 
temperature were determined for nuclei with neutron separation energies
less than about 2.75 MeV.  (It is not very useful to calculate decay rates
as a function of temperature for nuclei with $S_n>$2.75 MeV because, the
r-process path does not pass through these nuclei until well after freezeout.)
Of course, the neutron separation energies and capture rates are also affected by
nuclear single-particle excitations.  This becomes important 
especially along the r-process path where neutron separation energies are low.  
One may expect an increase in both the neutron emission rate as well as the
capture rate for excited-state nuclei though capture and photoemission are in
equilibrium in the classical r-process.  However, single-neutron excitations cannot
go above the neutron separation energy.  
For this reason,
single-neutron excitations are kept realistically low (though multiple-particle excitations may
be higher).  Indeed, even at r-process temperatures, one does not expect a significant
population of states above the neutron separation energy.

While the network calculation proceeds to Z=100, the mass formula used in this
work has shell energies up to N=157.  Therefore, neutron-rich nuclei with 
N$>$157 cannot be quantified, which effectively limits proton number Z to 
about 93.  
However, this is well past the A=195 peak in the r-process abundance 
distribution, and the effects of not including excited state rates for these
very heavy nuclei have little effect on the abundances of the lighter nuclei.
Also, it was found that the
abundances of fissile nuclei along the path are very low at freezeout, so that
the inclusion of fission cycling was ignored in this 
r-process model.
\subsection{Excited State $\beta$-Decays in the Reaction Network}
The inclusion of excited state $\beta$-decays into the reaction network can
be computationally simple.  Functional forms of decay rates with temperature 
were used.  
This allows a fairly accurate evaluation of the
calculated rates of individual nuclei for temperatures between 0 and 6 
billion
K, while still maintaining the speed of the network calculation.  
Sixteen regions of the isotopic chart were found for 
which decay rates 
of nuclei in these regions have similar dependences on temperature.  These 
regions are shown in Table \ref{temp_fits}.  
Nuclei represented in these figures have neutron separation energies
S$_n$ ranging between 0 and about 2.75 MeV.  This sort of parametrization allows for a 
very reasonable first approximation of the change in decay rates over a large region of the
isotopic chart.  

The dependence on temperature for the above mentioned regions can be
conveyed with a set of parameters fit to the formula:
\begin{equation}
\label{eq1-5}
\frac{\lambda}{\lambda_0}=E\left({FT_9+G}\right)\left[{A-
\frac{B}{\exp{\left({C\left({T_9-D}\right)}\right)}+1}}\right]
+{\alpha}\exp{\left({\beta{T_9}+\gamma{T_9^2}}\right)}
\end{equation}
where $\lambda$ is the $\beta$-decay rate at temperature T$_9$ (defined to be
the temperature in 10$^9$ Kelvin), and the ground-state decay rate is $\lambda_0$.

The ratio in Equation \ref{eq1-5} must be unity at T$_9$=0, so several of the
parameters are dependent:
\begin{eqnarray}
\label{eq2-5}
G & = & \frac{1}{E}
\nonumber\\
B & = & \left({A-1}\right)\left({\exp{\left({-CD}\right)}+1}\right)
\end{eqnarray}
so the parameter set is reduced to eight in number.  Further, in using 
Equation
\ref{eq1-5} to fit the dependences of rates to temperature, two terms 
were necessary in the temperature 
region of interest.  In actuality, only one of the
two terms is present at a time.  So if $E$ is non-zero, then
$\alpha$ is zero and vice-versa.  Also, $\alpha$ is limited to one of
two values - 0 or 1 - leaving seven free parameters.
The parameters for each of the 
regions mentioned above, along with the nuclei included in each region, 
are given in Table \ref{temp_fits}.

Two of the parameters in Table \ref{temp_fits} are functions of the nuclear
proton number Z.  These are the values of A for region 8 and the value of
$\gamma$ for region 7a, listed in the table as $\tilde{A}(Z)$ and 
$\tilde{\gamma}(Z)$ respectively.  This is a convenient way of representing
these parameters over a very large isotopic region, as the functional form 
of 
the temperature dependence of decay rate changes slowly with Z.  These 
parameters are given by:
\begin{eqnarray}
\label{eq3-5}
\tilde{A}(Z) & = & 100.31-1.38Z
\nonumber\\
\tilde{\gamma}(Z) & = & 0.185-0.003Z
\end{eqnarray}
No odd-even effects are observed because Equation \ref{eq1-5}
is a ratio of rates, so any local odd-even effects would be minimized in
the division.  Variations in the parameters as a function of mass seem to be
explainable in terms of local shell effects.
\subsubsection{Uncertainty Associated With the Use of the Spherical Shell Model}
It is worth mentioning the use and applicability of the spherical shell model in this formulation.  One should not
only consider effects on decay rates, but on the overall r-process.  While it is difficult to compare to a more
elaborate model in this work without utilizing such a model, a qualitative assesment is discussed here.  Two 
things which are not necessarily mutually exclusive
must be considered.  The first is the accuracy of the decay rates used.  The second is the applicability of the 
spherical shell model to the nuclei studied. 
In particular, one can estimate effects on individual transitions when one considers  a Nilsson shell model with 
deformation.  For this reason, nuclei which are presumed undeformed are compared to those which are considered
to lie in regions of deformity in the isotopic chart (i.e., between closed shells) are discussed.   

In estimating the accuracy of the calculated decay rates, one considers uncertainties in calculations for 
ground-state nuclei; the uncertainty in the decay rates may be estimated using Figure \ref{FRDM_rates}.
The rms error in S  (Equation \ref{eq1-4}) for all nuclei studied in this work with known $\beta^-$ decay 
rates has been calculated.
The rms value S$_{rms}$=$\langle$S$^2\rangle^\frac{1}{2}$ has 
for all nuclei, even-even nuclei, odd-N nuclei, odd-Z nuclei, and odd-odd nuclei to be 0.767,
0.569, 0.805, 0.745, and 0.872 respectively for the ground state nuclei, indicating reasonable agreement as shown
in Figure \ref{FRDM_rates} for at least the nuclei with known decay rates including those in regions of presumed
deformity.   The accuracy improves with rate as the presumed level density increases.  For nuclei with $\lambda>$0.1 s$^{-1}$ the
value of S$_{rms}$ is 0.579 and decreases further to 0.467 for nuclei with $\lambda>$ 1 s$^{-1}$, indicating the improvement in the
accuracy of this formulation for the r-process, in which the decay rates are larger.  The agreement is also equally 
good for the few
known neutron-rich isomers, though some caution is warranted here as the number of known isomeric states is small.  As many
of the nuclei studied were extremely neutron rich, the uncertainty in calculations is expected to be reasonable for those nuclei
along the r-process path.  In fact, the uncertainty is expected to decrease with single-particle level density.  While reasonable,
these errors may likely be reduced with a more realistic model.

In the case of non-deformed nuclei, the spherical shell model is believed to be a reasonable estimate of the
decay rates.  This is extremely important in the case of the r-process as the waiting points are localized 
about areas of lowest deformation; most of the r-process abundance is confined to these regions of the isotopic chart.
Large uncertainties in the decay rates through the waiting points can result in drastic changes in the final
distribution.  For example, consider the A$\sim$130 abundance peak.  If the decays rates of the r-process
nuclei associated with this peak were much higher, then an increased flow through this peak would result, resulting
in a much lower abundance of nuclei in this peak and a very large enhancement of the rare earth nuclei 
(130$<$A$<$190).

On the other hand, the portion of the r-process path thought to be responsible for the production of the rare-earth elements
passes through a region of the isotopic chart associated with possibly large deformations.  The overall effects on the final r-process 
abundance distribution due to changes in the decay rates of these nuclei is expected to be small as their contribution to the total
abundance is small.  In the case of these possibly deformed nuclei, single-particle levels
in the Nilsson model may shift by as much as 0.5 MeV from those of the spherical shell model in cases of extreme deformation.
However, single-particle levels are shifted in both the positive direction and the negative direction for protons and neutrons.
Qualitatively, the net effect is to break degeneracies in closed shells while ``spreading out'' high localized single-particle level densities
associated with closed shells.  

Consider, for example, Figures \ref{exc_func} and \ref{exc_func_ff} which show the calculated GT and first-forbidden 
strength functions for the $^{150}$Ba nucleus for various excitations.  If a more realistic structure model was used, the 
net result would be a spreading of the widths of the peaks of these functions as both the daughter and parent states would undergo 
shifts in both the positive and negative energies.  While some of the strength function would shift to higher transition energies (more
negative energies on the figure) resulting in contributions to higher decay rates, some would also shift to positive energies resulting in 
contributions to lower decay rates (and possibly - though unlikely - to states that cannot decay).   By shifting the transition 
energy by an extreme amount of 0.5 MeV for a GT transition in Equation \ref{eq10}, it is estimated that
the transition rates may change by as much as a factor of two.  However, it must be carefully noted that this is for a shift of the entire GT
transition strength in one direction for an extreme amount for only the most deformed nuclei.  Thus, this factor is an upper limit for 
very deformed nuclei, and likely the error due to using a more accurate model is found using the rms values in S as discussed above.  
Certainly future work may
concentrate on more realistic model calculations, though it will be seen from the next section that this may not be necessary as the 
results of the r-process calculations are more heavily dependent on uncertainties in the hydrodynamic conditions. 
\section{Results of the Network Calculation}
\label{network_results}
The results from several hydrodynamic parameter sets, as well as electron 
fraction
parameter values $Y_e$, were examined.  
These parameter sets are shown in Table \ref{net_parms}.  For each parameter set, the
core mass in solar masses, core radius, neutrino luminosity, initial
electron fraction, and whether or not $\beta$-delayed neutron emission
is included are listed.  Using these parameters and the calculations of reference 
\cite{otsuki00}
the dynamic timescale and the entropy in the
expansion are constrained.  Though still in agreement with current predictions,
the dynamic timescales in 
these calculations are shorter than average.   However, the entropy is 
lower, and no artificial increase in the entropy (as is often 
assumed) was required \cite{woosley94}.

\begin{figure}
\includegraphics[width=15cm]{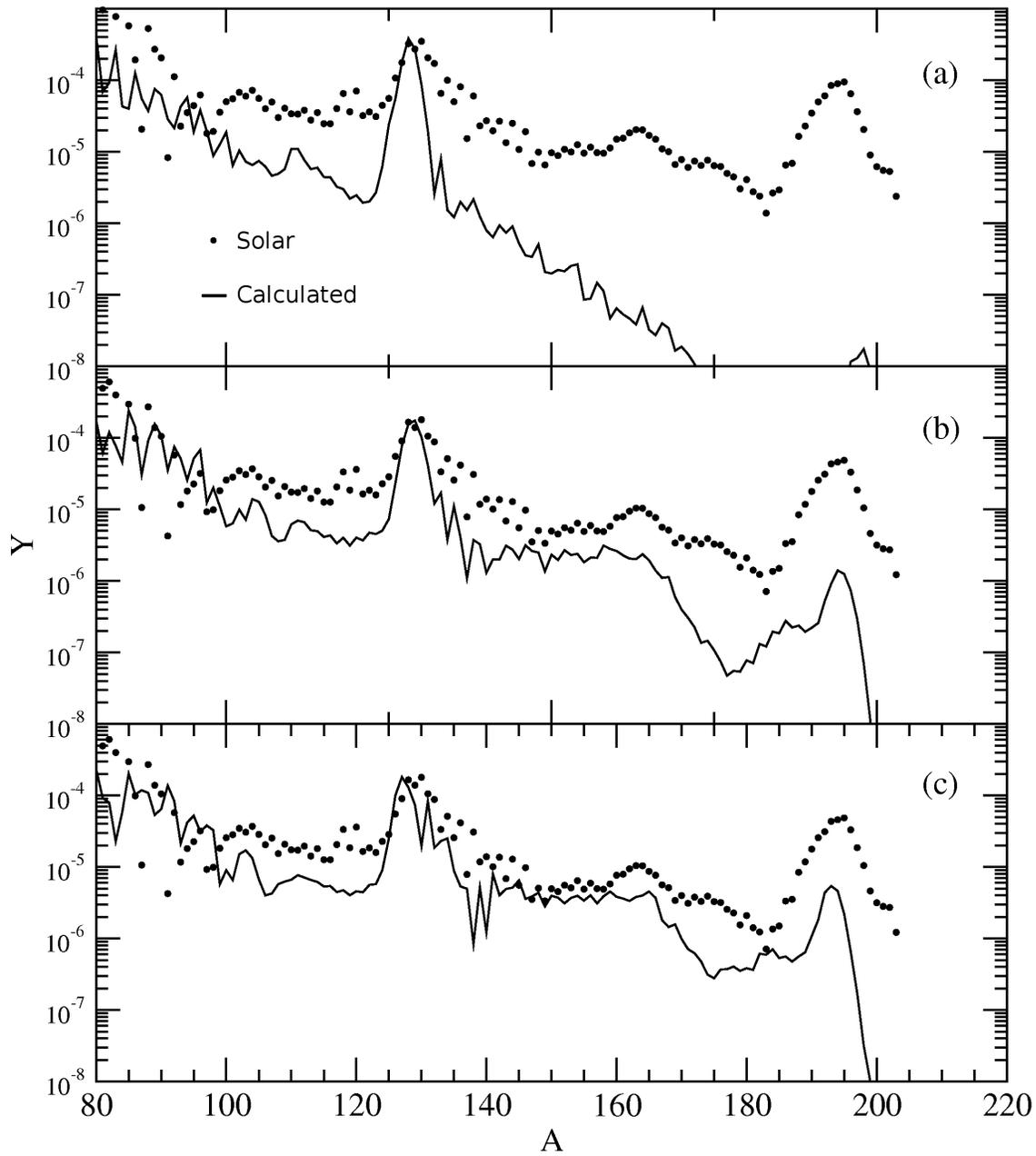}
\caption{\label{hyd_res}Final r-process abundance distributions for models A,
B, and C; shown in Figures (a), (b), and (c) respectively.  The solar r-process
abundance distribution (scaled to the figures) is given by the dots.
These models do not include $\beta$-delayed
neutron emission, and only include $\beta$-decays of ground-state 
nuclei.}
\end{figure}

Each simulation is run until several seconds beyond freezeout.  While this time is sufficient
to gauge the gross features of the r-process abundance distribution, a longer simulation may
have resulted in more post-processing, allowing for smoother abundance distributions.  
Figure \ref{hyd_res} shows that model A underproduces the A$\sim$195 peak by 
a large
amount.  This is due primarily to the fact that the lower entropy in
model A results in a very low neutron-to-seed ratio.  Models B and C were 
chosen as
intermediate points in the entropy-timescale phase space.  Both produce a 
more acceptable
r-process abundance distribution, although the A$\sim$195 peak is
still underproduced.  For comparison, the solar r-process 
abundance distribution is displayed in the figure.  One notes some residual even-odd
effects in the calculated distribution as the effects of smoothing may not be
complete, though the gross features of the distribution are noted.

\begin{figure}
\includegraphics[width=15cm]{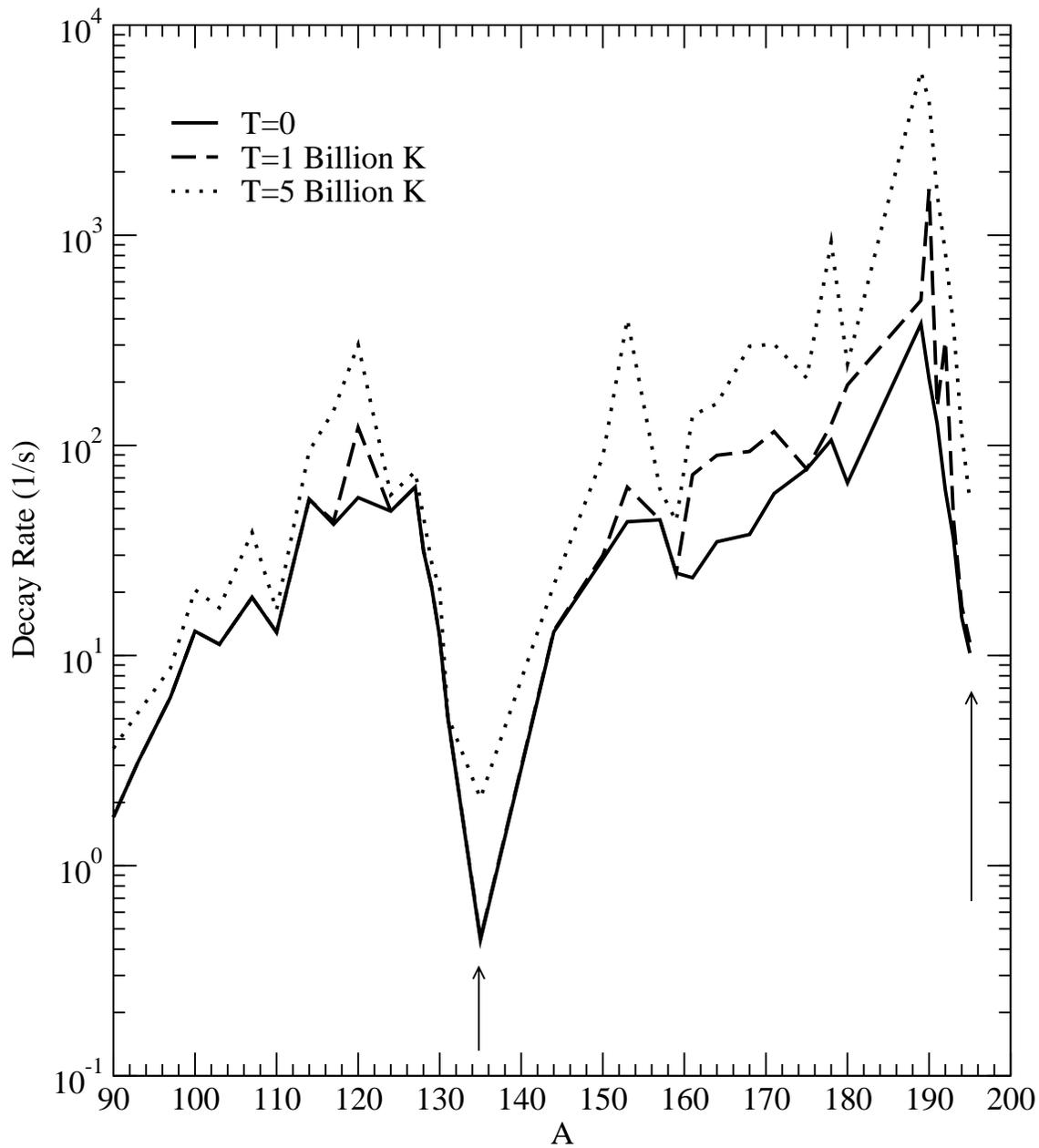}
\caption{\label{rel_rates}Calculated effective $\beta$-decay rates of
r-process nuclei used as a function of their mass.  The nuclei represented
in this figure fall along a line of neutron separation energy S$_n$=2.5 MeV.
The arrows indicate neutron closed shells at N=82 and 126.}
\end{figure}

\begin{figure}
\includegraphics[width=15cm]{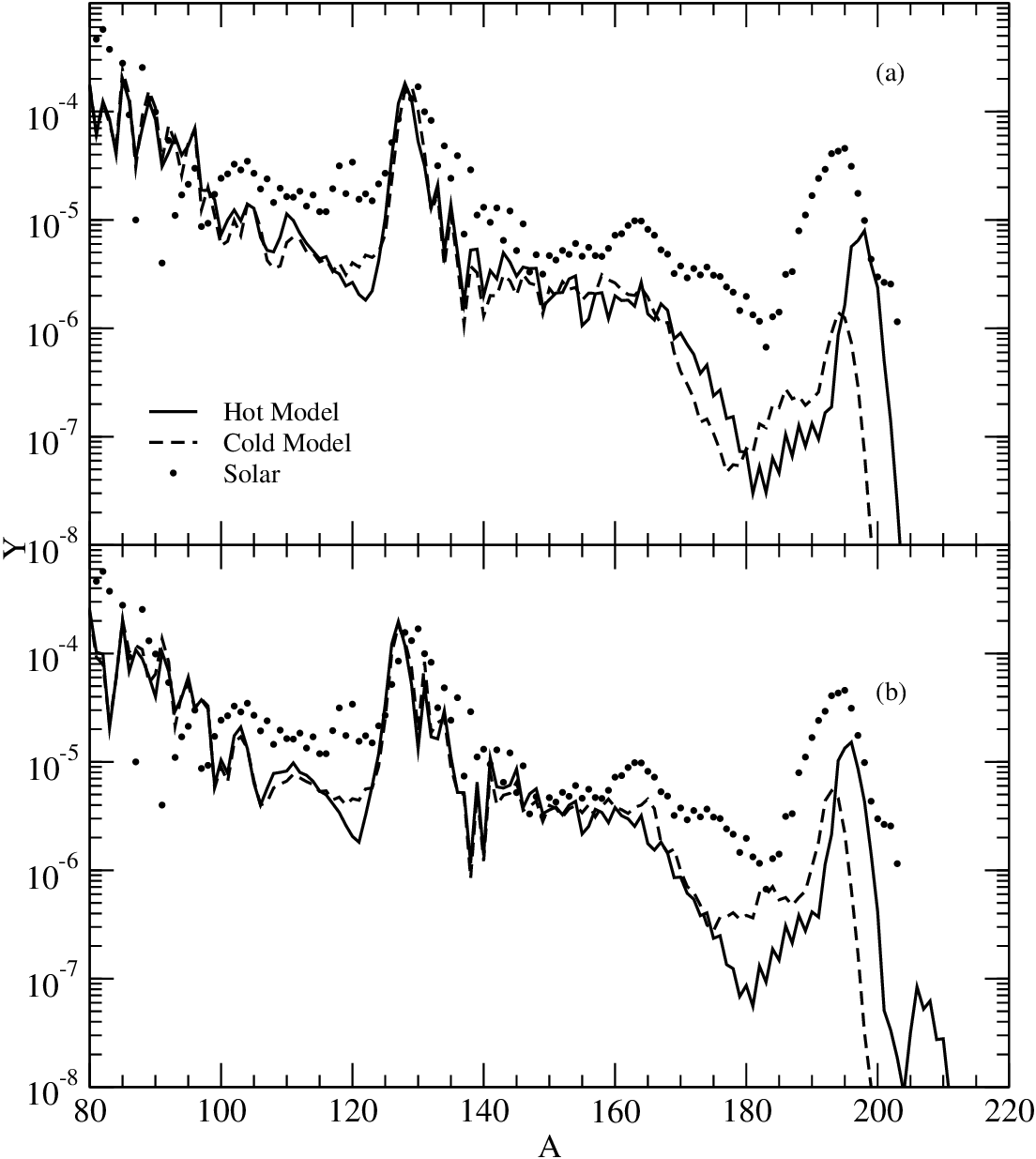}
\caption{\label{q_comp}Comparison of r-process freezeout abundance
distributions for models B and C in plots (a) and (b)
respectively.  Solid lines and dashed curves display the calculated results 
with (hot model) and without (cold model) excited-state $\beta$-decays
respectively.  The solar distribution is also shown by the dots.}
\end{figure}

Results from models B and C are displayed in Figure \ref{q_comp} 
for both the hot (i.e., including excited-state decays, solid line) and cold 
(i.e., not including excited state decays, dashed line)  models.  The 
A$\sim$195 peak is
most profoundly affected, along with the nuclei just below this
mass region.  From the relationship between the decay rates and 
temperature
(Equation \ref{eq1-5} and Table \ref{temp_fits}), it can be seen that decay 
rates of the
nuclei in the region just below the A$\sim$195 peak (and - to a lesser
extent - the region just below the A$\sim$130 peak) are quite sensitive to
changes in temperature even at low temperatures.  This is expected due to 
the high level densities of these nuclei (lying just below the N=126 and N=82 
closed shells), as discussed in \S\ref{grosstheory}.  Shell quenching has not been
included in this calculation, though the effect is noted, and no conclusions can be
drawn from this study to evaluate the effects of quenced closed shells far from 
stability.
Other rates, however, are not as sensitive to temperature changes at low
temperatures and, as the r-process progresses, these rates would drop
to their ground-state values before those of
the nuclei in the regions below the abundance peaks.  The decay rates of 
nuclei in this region would increase relative to those of nuclei in other
regions of the path, selectively depleting the abundances of nuclei in
this region.  This 
effect is displayed in Figure \ref{rel_rates}, in
which nuclei in these two regions have large changes
in decay rate as T$_9$ increases a small amount.

The lowering of the electron fraction in models F and G is a physically 
acceptable assumption given that the
electron neutrinos and anti-neutrinos captured on nucleons before the 
beginning 
of the $\alpha$-process
can alter the neutron excess $\eta$ via neutrino interactions on the
free nucleons in NSE.  In a steady-state, the electron fraction is 
\cite{fuller95}:
\begin{equation}
Y_e\approx\left({1+\frac{\langle{E_{\bar{\nu}_e}}\rangle}
{\langle{E_{{\nu}_e}}\rangle}}\right)^{-1}
\end{equation}
Given the neutrino energies assumed in the nucleosynthesis code, then
a value as low as $Y_e$=0.35 may not be unexpected, though higher values have
also been used.

\begin{figure}
\includegraphics[width=15cm]{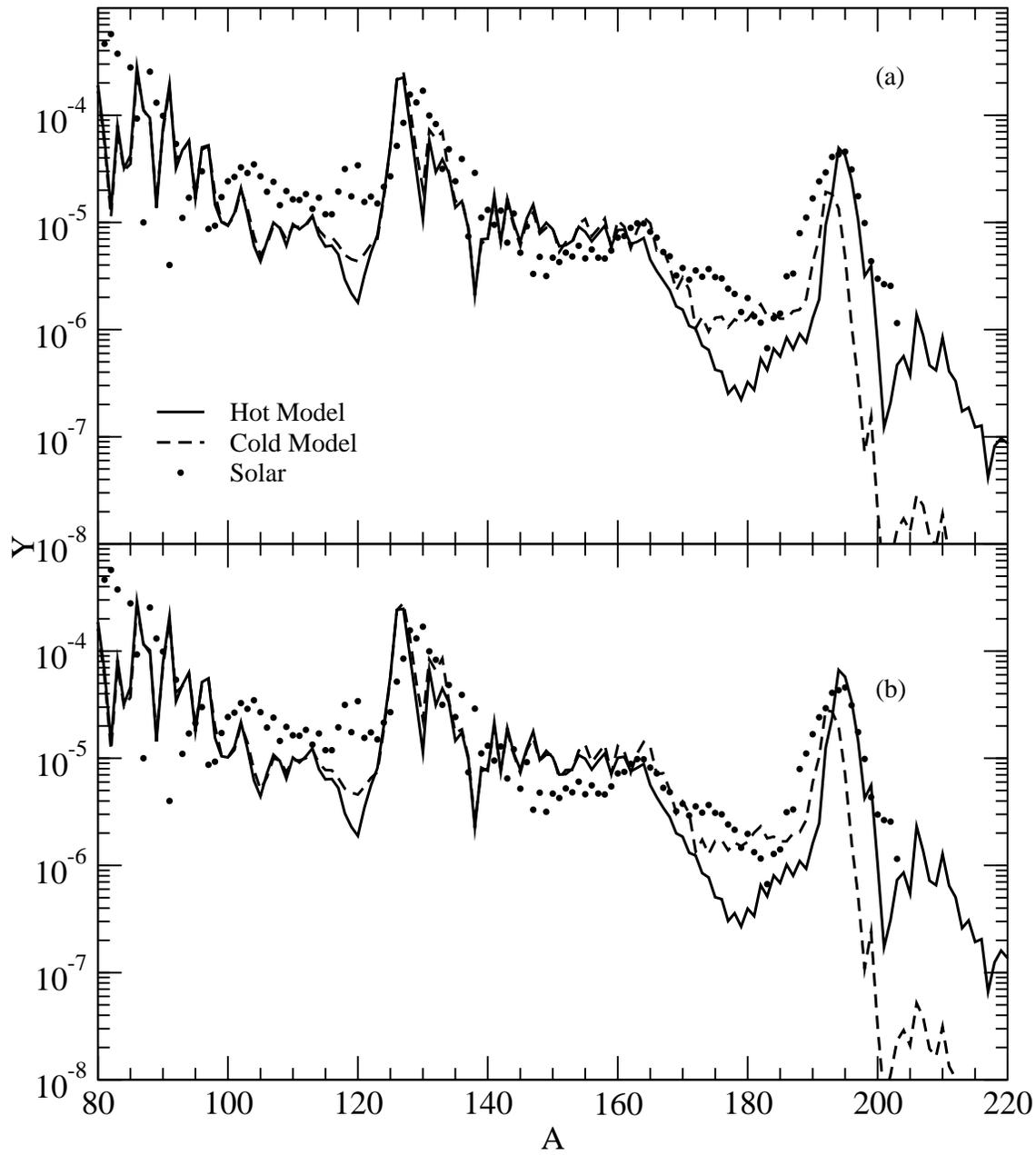}
\caption{\label{q_comp2}Comparison of r-process freezeout abundance
distributions for models F and G in plots (a) and (b),
respectively.  Solid lines and dashed curves display the calculated results 
with (hot model) and without (cold model) excited-state $\beta$-decays
respectively.  The solar distribution is also shown by the dots.}
\end{figure}

The abundance distribution results of models F and G are shown in Figure
\ref{q_comp2}.  The A$\sim$195 
peaks are more pronounced - closer to acceptable values - in both cases as compared to
models B and C.  Even more, 
abundances of nuclei with heavier masses are also increased, more closely 
matching the solar system abundance distribution in this region.  As in 
previous models,
the effect of excited state $\beta$-decays is still quite evident in both
models, especially for the heavy nuclei with A$>$195.  The ratios of peak 
abundances for these models are shown in 
Table \ref{relabs}.  Both models have peak abundance ratios close
to that of the solar system.  Also, in both models, the 
A$\sim$195 peak is shifted closer in position to the solar system abundance
peak.  The slight decrease in Y$_e$, while still maintaining physically
realistic values, combined with the use of excited-state $\beta$-decays,
produces an r-process calculation with results that match
the solar system distribution, with the exception of nuclei in the
A$\sim$180 region, which seem to exhibit a deficiency in abundance.

Figures \ref{q_comp} and \ref{q_comp2} suggest that the inclusion of excited-state
$\beta$-decays results in a net shift of abundance from the region of  nuclei just below the
A$\sim$195 abundance peak into this abundance peak.  Certainly, one notes that the more
rapid decay rates in the region just below the abundance peaks results in a faster progression
of the r-process through these nuclei.  However, one should also note that the integrated abundances
for nuclei with 180$<$A$<$200 is higher in the hot model by than in the cold model by 29\%, indicating that
overall abundance is increased in both the abundance peak and the region below this peak.  This suggests that
the increase in abundance in the A$\sim$195 abundance peak is only partially due to an increase in the decay rates
of the nuclei associated with the region just below this peak.  One must note that the overall abundance in the 
180$<$A$<$200 region increases due to a flow of abundance into this region from the lower masses.  This 
increased abundance can come from minute changes in abundance in the A$\sim$130 peak and from the
mass region just below this peak.  In the semi-log representation of Figures \ref{q_comp} and \ref{q_comp2}, it is seen
that even small fractional changes in the abundance of the A$\sim$130 peak - with nearly an order of magnitude more 
abundance than the A$\sim$195 peak - can result in sizable changes in the 180$<$A$<$200 mass region.  A similar 
comparison can be made between the abundance region just below the A$\sim$130 peak and the A$\sim$195 peak.

This brings up an interesting point regarding the net effect of rate increases with temperature in the r-process.  
Not surprisingly, figure \ref{rel_rates} shows that decay rates for the closed-shell nuclei are only affected at high
temperatures (corresponding to early stages in the r-process).  Very early in the r-process, flow through the A$\sim$130
mass region increases relative to the remaining nulcei along the r-process path, populating the rare earth region.  
However, as temperature drops before freezeout, the
relative flow through this mass region decreases, and the relative flow through the rare earth region is still higher, populating
nuclei in the A$\sim$195 mass region.  As the A$\sim$195 mass region is not populated until later (and cooler) in the r-process,
the effects of excited-state decays are minimal.  The net result is that as the r-process progresses to higher mass, the relative
decay rates drop on average first for the A$\sim$130 abundance peak and then for the rare earth region.  Abundances of
the r-process progenitor nuclei are affected early on by increased rates of the A$\sim$130 nuclei and later on by those of
the rare earth nuclei.  Of course, the net effect is dependent on the passage of the r-process path through many nuclei, so the 
network calculation is employed as a useful tool.

\begin{figure}
\includegraphics[width=15cm]{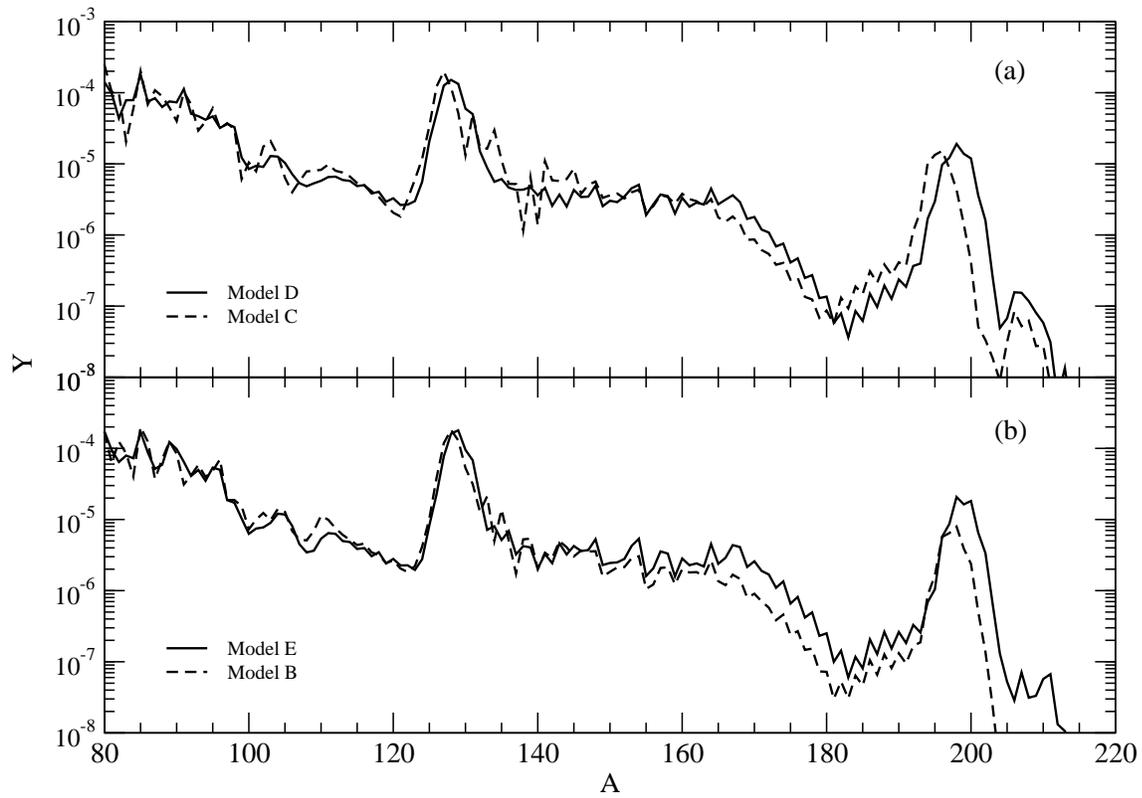}
\caption{\label{mod_s_comp}Freezeout r-process abundance distributions for 
models D and C, as well as models E and B in plots (a) and (b) respectively.  Both models include 
excited-state $\beta$-decays, while models D and E  included $\beta$-delayed
neutron emission.
}
\end{figure}

\begin{figure}
\includegraphics[width=15cm]{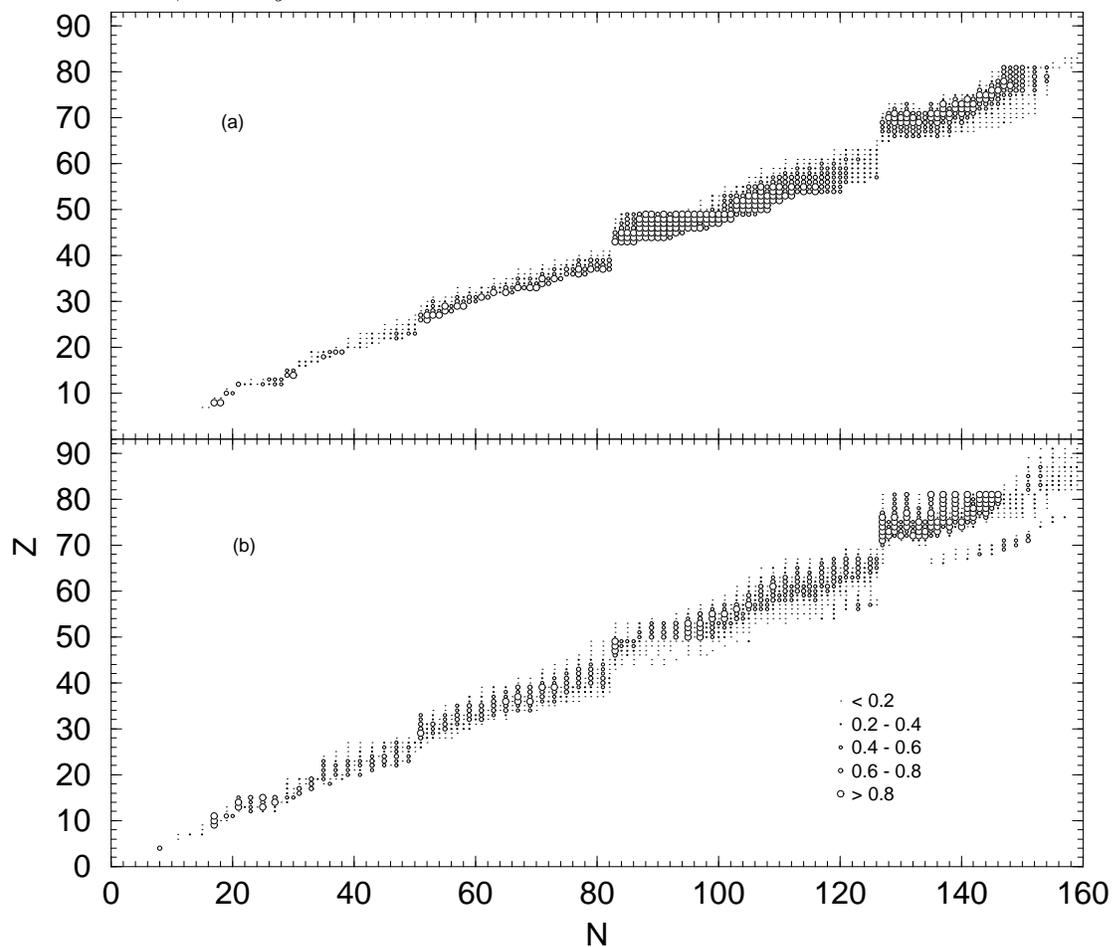}
\caption{\label{emissions}Double and single delayed neutron emission probabilities shown
in Figures (a) and (b) respectively.  It can be seen that the nuclei
just above the A=130 mass region have higher probabilities for two-neutron
emission following $\beta$-decay.}
\end{figure}

Models D and
E are identical to models C and B respectively, except for the fact that 
$\beta$-delayed neutron
emission reactions are included in models D and E.  The final freezeout 
abundance for model D is compared to that of model C, and the 
final freezeout abundance of model E is compared to that of B in
Figure \ref{mod_s_comp}.  The distribution
seems to be shifted slightly to heavier mass when $\beta$-delayed 
neutron emission is included.  While this may be surprising, just after 
freezeout, as nuclei
begin to $\beta$-decay back to stability, $\beta$-delayed neutrons 
become available for capture by all nuclei.  The 
two-neutron emission
probability as shown in Figure \ref{emissions} is higher in the 125$<$A$<$180 region than
in the region surrounding the A$\sim$195 peak, while the single-neutron
emission probability is slightly higher in the A$\sim$195 region, resulting
in a larger net loss of neutrons in the lower mass nuclei. 
Furthermore, the heavier nuclei tend to have higher neutron capture 
cross-sections
\cite{rolfs88}.  The overall result is that the masses of the 
heavy nuclei are increased by a few
mass units.  Figure \ref{mod_s_comp} shows that the mass shift of the
abundance peak is roughly two mass units, 
indicating that nuclei in the A$\sim$195 mass region 
have captured at least two net neutrons while decaying back to stability.  
This necessitates further study regarding the dynamical nature of the r-process after
freezeout \cite{surman97}; the r-process path continues to evolve to its
final distribution
even after the neutron abundance has dropped several orders of magnitude.

\section{Conclusion}
\label{conclusion}
This work provides a study of the effects of excited state 
$\beta$-decays on the r-process.  A preliminary method was used to evaluate the
possible effects of 
$\beta$-decay rates of excited-state nuclei.  
Though the accuracy of the model is
limited by the knowledge of single particle levels, an approximate treatment allows one
to gauge the magnitude of effects on the r-process and provide impetus for further study.  An 
empirical calculation was employed to find single-particle levels, and 
quantum numbers were deduced based on the level proximity to those of the 
spherical shell model.
Although minor effects were found to result from inclusion of the
excited state decays, there are also other possible effects 
that the inclusion of excited state nuclei may have on the r-process.  
One can imagine that if the excitation is due to the promotion of neutrons
to higher-lying single-particle orbitals, then the photoneutron Q-value will 
decrease, and the ($\gamma$,n) reaction rate may increase.  Thus the effect 
of an 
increased rate might be to shift 
the r-process path closer to $\beta$-stability.
  
All models used in this calculation do a reasonable qualitative job of 
reproducing the solar abundance distribution for the mass region
$80\le{A}\le{130}$.  The fact that the abundance at low mass is roughly 
independent of the type of model used (hot or cold) is an indicator that the 
nuclei in this region have decay rates not as heavily dependent on 
temperature as some of the higher mass nuclei.  
In the more massive nuclei, the sensitivity of the
decay rates on temperature resulted in a more pronounced shift in the path
as the r-process evolves through freezeout.

As mentioned previously, the dynamical treatment of
the r-process is an important factor here in that the path continues to
evolve even during freezeout, a result of $\beta$-delayed neutron
emission.  With the mass formula used in this evaluation, it was found that 
the low mass nuclei have a higher probability
of emitting two neutrons during $\beta$-decay than the higher mass nuclei, 
which have a higher probability of emitting a single neutron during 
$\beta$-decay.  These available 
neutrons are recaptured, with the cross
section roughly increasing with mass.  The net result is  a slight shift in 
the A$\sim$195 abundance peak.

Despite the ability to predict a more reasonable abundance of the A$\sim$190 nuclei, there is
still a discrepancy between the predicted 
abundance distribution in the rare earth region and that of the solar system.
The abundances of the rare earth elements (A$\sim$165) are predicted to
be lower 
in abundance by about an order of magnitude than the A$\sim$130 peak, in
fair agreement with that of the solar system.  This corresponds to the
argument of the authors of reference \cite{surman97}, who state that the rare-earth region
is a robust feature of any dynamical calculation including post-production
of the r-process progenitors.  However, nuclei with 130$<$A$<$160 are 
overproduced slightly, removing 
the effect of the rare earth region being manifest as a peak, hence the
appearance of the abundance distributions in the figure.    

It is obvious that no shell quenching has been employed in this preliminary model, as can be seen
by the dip in the A$\sim$180 abundance.  
While the abundance of the A$\sim$180 nuclei relative to
the A$\sim$130 peak is similar to that of the solar system, the width of this
dip is greater than that of the solar system abundance distribution.
It has been mentioned 
\cite{kratz93} that the underproduction of nuclei in this region might vanish if the quenching of shell closures
for the very neutron-rich nuclei in this mass region is properly included.  

Because of the large number of nuclei involved, the improved semi-gross theory has been 
used to globally calculate
decay rates.  It is understood that the initial model represents a first attempt to gauge the effects of nuclear 
$\beta$-decays in the hot environment of the r-process, and further study is warranted.  
In particular, a more accurate global calculation of decay rates is desired.  Currently, the measurements of
$\beta$-decay rates are limited to either ground-state nuclei or long-lived isomers.  However, with the advent of 
large neutron flux devices\cite{petrasso96}, 
the measurement of the GT strength functions of nuclei in excited states from may become
feasible within the next decade, allowing for experimental confirmation of the transition strengths of excited-state nuclei. 
\ack
The authors wish to acknowledge the helpful insight provided by K. Takahashi and
comments by B.A. Brown.
This work was funded by NSF grants PHY 9901241 and PHY9905241 and by the WMU Faculty Research
and Creative Activities Support Fund (FRACASF) grant \#06-005.

\newpage
\clearpage
\begin{table}
\caption{\label{terms} Terms Used in Gross Theory Presented in This Paper} 
\begin{indented}
\item[]\begin{tabular}{@{}lll}
\br
Term & Description & Equations \\
\mr
N$_1$ & Number of Parent Nucleon States & 1 \\
W & Weighting Function Describing & 1,5\\
 & Availability of Daughter Nucleon States & \\
$\varepsilon$ & Energy of a Nucleon State & 1-4,4,17,18,20\\
D$_\Omega$ & Transition Probability Function & 1,7,9,10\\
$\varepsilon_0$ & Lowest Energy of a Parent Nucleon & 1\\
 &Energetically Able to Decay &  \\
$\varepsilon_1$ & Highest Energy of an occupied Parent Nucleon State & 
1-3 \\
n$_3$ & Number of Daughter Nucleons Collected at the & 3 \\
 & Highest Daughter Level Due to the Pairing Force &  \\
2$\Delta$ & Width of the Pairing Gap & 2,3 \\
\br
\end{tabular}
\end{indented}
\end{table}

\begin{table}
\caption{\label{termsb}Terms Used in the Single-Particle
Rate Calculations Presented in this Paper}
\begin{indented}
\item[]\begin{tabular}{@{}cll}
\br
Term & Description & Equations\\
\mr
n$_i$ & Number of Parent or Daughter Nucleons in Level i & 4,5,11,12,21 \\
$\omega$ & Spherical Shell Model Eigenstate Mixing & 6,7,11,12,20 \\ 
~& Coefficient for Mapping to Leves & ~ \\
~ & in the FRDM & ~ \\
$\Lambda_\Omega$ & Function for Describing Selection & 7,8,11,12,21 \\
~& Rules of a Transition $\Omega$ & ~\\
g$_\Omega$ & Multiplicity Spin-Averaging Factor for & 11 \\
~ & Transition Type $\Omega$ & ~ \\
$\varepsilon_k^0$ & Discrete Uncorrected Fermi Energy of & 15,16 \\
~ & the k$^{th}$ Level & ~ \\
d$_k$ & Standard Level Density About the k$^{th}$ Level & 15,16 \\
H($\varepsilon$) & Triangular Distribution Function Indicating & 14,17,20 \\
~ & the Strength of the Spherical Shell Model Eigenstates & ~ \\
q(A) & Width of the Triangular Function H($\varepsilon$) & 14,15 \\
G($\varepsilon$) & Sum of H($\varepsilon$) Over All Eigenstates & 17,18 \\
$\varepsilon^\pm$ & Integration Limits About G($\varepsilon$) and & 18-20 \\ 
~ & H($\varepsilon$) for Normalizing and Determining $\omega$. & ~ \\
\mr
\end{tabular}
\end{indented}
\end{table}

\begin{table}
\caption{\label{frdm_param}Parameters Used in the FRDM Nuclear Mass Model} 
\begin{indented}
\item[]\begin{tabular}{@{}cllc}
\br
Parameter & Description & Value & Units \\
\mr
$\Delta{M}_H$ & Hydrogen atom mass excess & 7.289 & MeV \\
$\Delta{M}_n$ & Neutron mass excess & 8.071 & MeV \\
$a_{el}$ & Electronic binding constant & $1.433\times{10}^{-5}$ & MeV \\
$K$ & Nuclear compressibility constant & 240 & MeV \\
$W$ & Wigner constant & 30 & MeV \\
$a_0$ & $A^0$ constant & 0.0 & MeV \\
$a_1$ & Volume-energy constant & 16.247 & MeV \\
$a_2$ & Surface energy constant & 22.92 & MeV \\
$a_3$ & Curvature energy constant & 0 & MeV \\
$J$ & Symmetry energy constant & 32.73 & MeV \\
$Q$ & Effective surface-stiffness constant & 29.21 & MeV \\
$c_a$ & Charge asymmetry constant & 0.436 & MeV \\
$I$ & Relative neutron excess & $\frac{N-Z}{A}$ & - \\
$\bar{\Delta}_n$ & Average neutron pairing gap &  $\frac{r_{mac}B_s}{N^{1/3}}$ & MeV \\
$\bar{\Delta}_p$ & Average proton pairing gap &  $\frac{r_{mac}B_s}{Z^{1/3}}$ & MeV \\
$\bar{\delta}_{np}$ & Average neutron-proton interaction energy &  $\frac{h}{B_sA^{2/3}}$ & MeV \\
$c_1$ & Coulomb energy constant & $\frac{3}{5}\frac{e^2}{r_0}$ & MeV \\      
$c_2$ & Volume redistribution energy constant & 
$\frac{1}{336}\left({\frac{1}{J}+\frac{18}{K}}\right)c_1^2$ & MeV \\
$c_4$ & Coulomb exchange constant & $\frac{5}{4}\left({\frac{3}{2\pi}}\right)^{2/3}c_1$ & MeV \\
$c_5$ & Surface redistribution energy constant & $\frac{1}{64Q}c_1^2$ & MeV \\
$f_0$ & Form factor correction constant & $-\frac{1}{8}\left(\frac{145}{48}\right)\frac{r_p^2e^2}{r_0^3}$
& MeV \\ 
$r_0$ &  Nuclear radius constant & 1.16 & fm \\
$r_p$ & Proton rms radius & 0.80 & fm \\
$r_{mac}$ & Average pairing gap constant & 4.80 & MeV \\
$h$ & Neutron-proton interaction constant & 6.6 & MeV \\
$a_{den}$ & Diffuseness of Yukawa function & 0.70 & fm\\
$r_{mic}$ & LN pairing constant & 3.2 & MeV\\
\mr
\end{tabular}
\end{indented}
\end{table}

\begin{table}
\caption{\label{temp_fits}Parameters Used for Fitting Decay Temperature Dependence}
\begin{indented}
\item[]\begin{tabular}{@{}ccccccccccc}
\br
~&~&~&\centre{8}{Parameter}\\
\centre{2}{Region}&Label&\crule{8}\\
~&~&~&A&C&D&E&F&$\alpha$&$\beta$&$\gamma$\\
\mr
\centre{2}{67$\le$N$\le$75} & 1 & - & - & - & 0 & - & 1 & 0 & 
0.03 \\
\ms
\centre{2}{76$\le$N$\le$77} & 2 & 2 & 2 & 2.5 & 0.3 & 1 & 
0 & - & - \\
\ms
\centre{2}{N=78} & 3 & 3 & 9 & 0.25 & 0.095 & 1 & 0 & - & - \\
\ms
\centre{2}{N=79} & 4 & 1.25 & 5 & 0.5 & 0.2 & 1 & 0 & - & - \\
\ms
\centre{2}{80$\le$N$\le$81} & 5 & - & - & - & 0 & - & 1 & 0 &
0.018 \\
\ms
\centre{2}{N=82} & 6 & - & - & - & 0 & - & 1 & 0.032 & 0.045 \\
\mr
~ & 39$\le$Z$\le$42 & 7a & - & - & - & 0 & - & 1 & 0 & $\tilde{\gamma}(Z)$ \\
~ & 47$\le$Z$\le$51 & ~ & ~ & ~ & ~ & ~ & ~ & ~ & ~ & ~ \\ 
\ns
~&~&\crule{9}\\
\ns 
~ & 43$\le$Z$\le$46 & 7b & 9 & 1 & 2.5 & 0.2 & 1 & 0 & - & - \\
82$<$N$<$126 & 52$\le$Z$\le$54 & 7c & 12 & 1 & 4 & 
0.28 & 1 & 0 & - & - \\
~ & Z=55  & 7d & - & - & - & 0 & - & 1 & 0.08 & 0.025 \\
~ & Z$\ge$56 & 7e & 12 & 8 & 0.5 & 0.04 & 1 & 0 & - & - \\
\mr
\centre{2}{N=126} & 8 & $\tilde{A}$ & 4 & 1.2 & 0.3 & 1 & 0 & - & - \\
\mr
~ & 60$\le$Z$\le$63 & 9a & 5 & 3 & 0.75 & 0.25 & 1 & 0 & - & - \\
~ & 64$\le$Z$\le$66 & 9b & 20 & 0.88 & 7.3 & 1 & 0 & 0 & - & - \\
N$>$126 & 68$\le$Z$\le$71 & ~ & ~ & ~ & ~ & ~ & ~ & ~ & ~ & ~ \\
\ns
~&~&\crule{9}\\
\ns
~ & Z=67 & 9c & 1.5 & 5 & 0.25 & 0.4 & 1 & 0 & - & - \\
~ & 72$\le$Z$\le$74 & 9d & - & - & - & 0 & - & 1 & 0.3 & 0.025 \\
\br
\end{tabular}
\end{indented}
\end{table}

\begin{table}
\caption{\label{net_parms}Parameters Used in the Supernova Model}
\begin{indented}
\item[]\begin{tabular}{@{}cccccccc}
\br
Model & M/M$_\odot$   & R(km)   & L$_\nu$ (10$^{52}$erg s$^{-1}$) & Y$_e$  & ($\beta$,n)? & $\tau_{dyn}$ (s) & S (k)\\
\mr
A & 1.7 & 10 & 0.7 & 0.4 & N & 0.017 & 110 \\
B & 2 & 10 & 0.7 & 0.4 & N & 0.014 & 151 \\
C & 2 & 10 & 0.5 & 0.4 & N & 0.021 & 162 \\
D & 2 & 10 & 0.5 & 0.4 & Y & 0.021 & 162 \\
E & 2 & 10 & 0.7 & 0.4 & Y & 0.014 & 151 \\
F & 2 & 10 & 0.5 & 0.37 & Y & 0.021 & 162\\
G & 2 & 10 & 0.5 & 0.35 & Y & 0.021 & 162 \\
\br
\end{tabular}
\end{indented}
\end{table} 

\begin{table}
\caption{\label{relabs}Relative Abundances of the A$\sim$195 and A$\sim$130
Peaks in the r-Process Calculations (Hot Models) }
\begin{indented}
\item[]\begin{tabular}{@{}ccccc}
\br
Model & A$_{max,195}$   & A$_{max,130}$   &
Y(A$_{max,195}$)/Y(A$_{max,130}$) &
$\sum{A{\sim}195}/\sum{A{\sim}130}$\\
\mr
B & 198 & 128 & 0.04 & 0.05\\
C & 196 & 127 & 0.07 & 0.09\\
D & 198 & 128 & 0.13 & 0.12\\
E & 198 & 129 & 0.12 & 0.13\\
F & 194 & 127 & 0.22 & 0.23\\
G & 194 & 127 & 0.27 & 0.27\\
Solar & 195 & 130 & 0.27 & 0.29\\
\br
\end{tabular}
\end{indented}
\end{table} 

\newpage
\clearpage
\newpage
\newpage
\clearpage
\newpage
\newpage
\newpage
\newpage
\newpage
\clearpage
\newpage
\newpage
\newpage
\newpage
\newpage
\newpage
\newpage
\newpage
\newpage
\newpage
\newpage
\newpage
\newpage
\newpage
\end{document}